\documentclass[graybox, nosecnum]{svmult}

\usepackage{mathptmx}       
\usepackage{helvet}         
\usepackage{courier}        
\usepackage{type1cm}        
\usepackage{makeidx}         
\usepackage{graphicx}        
\usepackage{multicol}        
\usepackage[bottom]{footmisc}
\usepackage{hyperref}        
\usepackage{soul}            

\usepackage{aas_macros}

\usepackage{natbib}
\usepackage[svgnames]{xcolor}
\usepackage{soul}
\usepackage{acronym}
\usepackage{xspace}
\usepackage{amsmath,amssymb,amsfonts,amscd}

\newcommand{\msol}{M_{\odot}}

\newcommand{\Msol}{M_{\odot}}

\newcommand{\ie}{i.e.\xspace}

\newcommand{\eg}{e.g.\xspace}
\newcommand{\viz}{viz.\xspace}
\newcommand{\wrt}{w.r.t.\xspace}

\newcommand{\erg}{\textrm{erg}}

\newcommand{\AZNucleus}[3]{{}^{#1}_{#2}\mathrm{#3}}

\newcommand{\figref}[1]{Fig.\,\ref{#1}}

\newcommand{\Alfven}{Alfv{\'e}n\xspace}

\makeindex             

\begin{document}
\title*{Nucleosynthesis in jet-driven and jet-associated supernovae}
\titlerunning{Jet supernovae} 
\author{Martin Obergaulinger$^1$ \thanks{corresponding author} and Moritz Reichert$^2$}
\institute{
$^1$ Departament d'Astonomia i Astrof\'{\i}sca, Universitat de Val\`encia, Edifici d'Investigatci\'{o} Jeroni Munyoz, C/Dr. Moliner, 50, E-46100 Burjassot (Val\`encia), Spain \\ \email{martin.obergaulinger@uv.es}
\\ 
$^2$ Departament d'Astonomia i Astrof\'{\i}sca, Universitat de Val\`encia, Edifici d'Investigatci\'{o} Jeroni Munyoz, C/Dr. Moliner, 50, E-46100 Burjassot (Val\`encia), Spain  \\ \email{moritz.reichert@uv.es}
}
%
%
\maketitle
\abstract{
  In contrast to regular core-collapse supernovae, explosions of
  rapidly rotating massive stars can develop jets, fast collimated
  outflows directed along the rotational axis.  Depending on the rate
  of rotation and the magnetic field strength before collapse as well as on
  possible mechanisms amplifying the magnetic field, such a core can
  explode magnetorotationally rather than via the standard supernova
  mechanism based on neutrino heating.  This scenario can explain the
  highest kinetic energies observed in the class of hypernovae.  On
  longer time scales, rotation and magnetic fields can play an
  important role in the engine of long gamma-ray burst powered by
  proto-magnetars or hyperaccreting black holes in collapsars.  Both
  classes of events are characterized by relativistic jets and winds
  driven by neutrinos or magnetic spin-down of the central objects.
  The nucleosynthesis in these events includes the production of Fe
  group elements, including a possibly enhanced synthesis of
  radioactive $^{56}$Ni leading to high peak luminosities.  Additionally,
  these events are, out of all stellar core-collapse events the ones
  most likely to allow for the formation of the heaviest nuclei via
  rapid neutron captures.  Increasingly sophisticated numerical
  simulations indicate that at least a limited r-process is possible,
  though it remains open how robust this result is against variations
  in the numerical methods and the initial conditions.  If so,
  supernovae with jets could contribute to the observed galactic
  chemical enrichment, in particular at early times before
  neutron-star mergers might be able to set in.
}


\section{Introduction}
\label{Sek:Intro}

The many different phases through which stars evolve set the stage for
several processes that produce and release almost all of the chemical
elements observed in the universe.  Of particular importance are
core-collapse supernova (CCSN) explosions \citep[for reviews, see, e.g.,][]{Janka_etal__2007__PRD__SN_theory,Janka__2012__ARNPS__ExplosionMechanismsofCore-CollapseSupernovae,Thielemann_et_al__2018__ssr__NucleosynthesisinSupernovae,Mueller__2020__LivingReviewsinComputationalAstrophysics__HydrodynamicsofCoreCollapseSupernovaeandTheirProgenitors}. They terminate the lives of
massive stars of more than about $8 \, \Msol$ after they have gone
through a series of burning stages in hydrostatic equilibrium.  When
nuclear reactions in the core of such a star cease to be net
exothermic and its mass grows to approximately the Chandrasekhar limit
of $M_\mathrm{Ch} \sim 1.4 \, \Msol$, gravitational instability sets
in and the core collapses.  When the central density exceeds the
nuclear saturation density, the collapse stops and the core turns into
a proto-neutron star (PNS) of a radius of at first a few tens of km.
Matter bounces back at its surface and a shock wave is launched from
there.  As the post-shock gas loses energy via neutrino radiation and photodissociation, the shock wave stalls inside the core rather than reaching the stellar surface.  It turns into a stalled accretion shock through which matter
continues to fall onto the PNS.

While most collapsing cores reach this common state, the further
evolution depends sensitively on the properties of the progenitor
stars \citep{Woosley_Heger_Weaver__2002__ReviewsofModernPhysics__The_evolution_and_explosion_of_massive_stars,Smartt__2009__araa__Progenitors_of_CCSNe,Langer__2012__ARAA__Presupernova_Evolution_of_Massive_Single_and_Binary_Stars}, in particular on how compact the core is, how fast it rotates,
and how strong its magnetic fields are.
All of these properties depend in turn on the initial conditions set
at the birth of the star such as the zero-age main-sequence mass, $M_{\mathrm{ZAMS}}$, the
metallicity, the initial rotational energy and magnetization of the
star as well as on many uncertain processes during stellar evolution
such as convection, small-scale magnetic fields, mass loss, binary
evolution.  These elements are not independent of each other.  Low
metallicity, \eg, causes star formation to favor more massive stars
and reduces the mass loss, which can make stars retain more angular
momentum when they collapse \citep[e.g.,][]{Brott2011}.

Increasingly sophisticated numerical simulations have demonstrated
that the revival of the stalled shock wave, leading to a successful
CCSN, is possible for stars in a wide range of masses above
$M_{\mathrm{ZAMS}} \gtrsim 8 \, \Msol$ without rapid rotation and
magnetic fields.  In what is regarded the standard CCSN mechanism,
neutrinos emitted in and around the hot PNS deposit energy in the
surrounding layers behind the shock wave \citep{Colgate1966,Arnett1966,Arnett1967}.  Non-spherical fluid flows
driven by convection or the standing accretion shock instability
(SASI) enhance the efficiency of neutrino heating.  The resulting
explosions take place with a delay of the order of a few 100 ms \wrt
core bounce. Their energies tend to be on the lower side of the
canonical CCSN explosion energy of $E_{\mathrm{exp}} \sim 10^{51} \,
\erg$, and the peak electromagnetic luminosities may reach those
observed in most CCSNe.  Under the influence of the hydrodynamic
instabilities, gas is ejected in clouds with a broad range of sizes
propagating in random directions.  The resulting distribution of
matter in position space and velocities is consistent with the
observations of many CCSNe and their remnants.

The onset of an explosion and the ejection of most layers of the star
greatly reduce the rate at which matter falls toward the PNS.
Typically the PNS reaches a mass well below the maximum value for
stability set by the nuclear equation of state (EOS).  In this case,
it turns into a young neutron star (NS).  During the transitory
Kelvin-Helmholtz phase, it gradually cools down by the emission of
neutrinos.  These neutrinos can drive a fast, hot wind in the
surrounding tenuous gas.

If neutrino heating is too weak to revive the stalled shock wave, mass
accretion may continue for much longer time, eventually causing the
the PNS to become gravitationally unstable and collapse to a black
hole (BH).  Further emission of neutrinos is suppressed.  Little mass,
if at all, is expelled and the result will be a failed CCSN.

While the majority of CCSNe follow the evolutionary path described
above, several classes of events remain difficult to explain this way,
in particular the most extreme ones in terms of explosion energy, peak
luminosity, or outflow speeds \citep{Woosley_Bloom__2006__araa__The_Supernova_Gamma-Ray_Burst_Connection}:
\begin{enumerate}
\item Hypernovae (HNe) are rare and very powerful CCSNe,
  releasing roughly ten times the kinetic energy of an ordinary
  supernova.  Their narrow spectral lines indicate collimated,
  relativistic outflows in addition to more spherical ejecta carrying
  the bulk of the energy and mass.  Typically belonging to spectral
  type Ic, they are associated to progenitors that have lost their
  outer (H and He) envelopes.
\item Superluminous SNe (SLSNe) emit electromagnetic luminosities far
  exceeding that of a standard CCSN and even that of thermonuclear
  type Ia supernovae by up to an order of magnitude.  Like HNe, they
  come from stripped envelope progenitors.
\item Several gamma-ray bursts (GRBs) with long duration of more than
  2 seconds as well as X-ray flashes (XRFs) show signatures of CCSNe
  several days to weeks after the initial high-energy transient.  The
  high-energy emission (GRBs, XRF) is generated by highly relativistic
  outflows accelerated by a central engine formed as the result of
  stellar core collapse.
\end{enumerate}

The boundaries between these classes are fuzzy.  The CCSNe directly
detected in the light curves of GRBs are classified as HNe, but many HNe are observed without a GRB and vice
versa.  In many cases, the distinction may be due to
observational biases.  An off-axis GRB, \eg, may be difficult to
detect, as may be a weak CCSN signature in a very distant GRB.
Nonetheless, there are enough events blurring the boundaries to
suspect a feature common to most of them, \viz jets, here understood
as collimated outflows of fast material.

Jets can be found in many astrophysical systems from protostars to
active galactic nuclei.  They are usually accelerated by a central
engine and ultimately powered by the gravitational energy which is liberated by
accretion or by the rotational energy extracted from the central
object. Although the exact mode by which this energy reservoir is
converted into the kinetic energy of the outflowing gas and hence also
its thermodynamic conditions depend on the specific system.  

In the case of stellar core collapse, the engines are usually assumed
to be a newly formed stellar-mass BH surrounded by an accretion disk
or a rapidly rotating and very strongly magnetized (P)NS, a
so-called (proto-)magnetar.  The outflows originate from gas
temporarily stored in the accretion disk or from the immediate
environment of the PNS and are accelerated by the magnetic field of
the disk/BH system or the PM or by asymmetric deposition of energy by
neutrinos. In any case, rapid rotation is essential for the engine: in one
case, it halts the radial fall of gas onto the BH and enables the
formation of the disk, in the other case, it serves as the source of
energy from which the kinetic energy of the jets is extracted.

The important role of rapid rotation restricts the
viability of these explosion mechanisms as most progenitor stars
rotate only slowly.  Even if formed with high rotational speeds,
massive stars suffer large losses of angular momentum due to their
intense radiatively driven winds.  Hence, these are rare
events, which, however, may release more mass or matter under more
extreme conditions than regular CCSNe.  While it is hard to make
precise statements about the rates of such events, the lower
opacities of gas at low metalicities reduce the stellar winds and,
thus the loss of mass and angular momentum.  Hence, the galaxies in
the early universe may have seen a larger fraction of extreme core
collapse events.

Besides these special classes, there may be a role for jets even in
ordinary CCSNe \citep{Soker_et_al__2013__AstronomischeNachrichten__Thejetfeedbackmechanism(JFM):Fromsupernovaetoclustersofgalaxies,Papish__2014__MonthlyNoticesoftheRoyalAstronomicalSociety__ExplodingCoreCollapseSupernovaebyJetsDrivenFeedbackMechanism}.  
{The turbulent dynamics of the post-bounce core can generate the conditions for jet formation if, possibly intermittently, the PNS acquires a strong magnetic field and a high angular momentum.  These jets would be weaker and slower than the ones discussed above and change their propagation direction randomly.}

All successful CCSNe are important production sites for heavy
elements \citep[e.g.,][]{Thielemann_et_al__2018__ssr__NucleosynthesisinSupernovae}.  Which processes are at work and which elements can be
formed very much depend on the precise dynamics, in particular the
electron fraction of the gas, its temperature and density, and the
speed at which it expands.  These parameters can be very different
between the classes of explosions listed above and even for gas that
is located in different regions of the same star.  Once the CCSN sets
in, explosive nucleosynthesis forms elements up to and beyond the Fe
group behind the shock wave.  Among the elements produced in this way,
several radioisotopes, in particular $\AZNucleus{56}{28}{Ni}$, are of
great observational interest as their later decay is largely
responsible for the photon emission during weeks and months after the
explosion.

The r-process as the most important processes for producing the
heaviest elements in stellar core collapse
relies on rapid captures of neutrons on seed nuclei and subsequent
beta decays.  It  operates most effectively under rather narrow
conditions, with high entropies and/or low electron fractions being most
conducive.  Early studies of CCSN nucleosynthesis had shown great
potential for the r-process in the neutrino-driven winds of ordinary
events.  However, later simulations with improved physics did not
confirm this possibility.  The crux of the matter is that reactions
with neutrinos not only transfer energy to the ejected gas, but also
alter its composition.  As a consequence, the ejecta would not be
neutron-rich enough to maintain an r-process.

Nowadays, the r-process is discussed in connection to stellar core
collapse almost exclusively for extreme explosions in which jets play
an important role \citep{Nomoto2017,Thielemann_et_al__2018__ssr__NucleosynthesisinSupernovae,Cowan__2021__ReviewsofModernPhysics__OriginoftheHeaviestElementstheRapidNeutronCaptureProcess}.  In particular when driven by magnetic fields, jets
offer the possibility to extract neutron-rich matter from close to the
centre (PNS or accretion disk) at high speeds without modifying the
electron fraction.  

The conditions for the r-process have to be evaluated for each possible site by a
combination of methods.  Large-scale numerical simulations of the
explosion reveals the dynamics.  Huge efforts are
necessary to reliably predict the thermodynamic conditions:
simulations have to be performed in three spatial dimensions, for many
dynamical timescales, and with detailed, and computationally costly,
neutrino physics. These models subsequently serve as an input for calculations of
the detailed nucleosynthesis yields using very large nuclear reaction network
codes.  

Testing these theoretical results against observations is difficult.
First of all, the events are rare, in particular in the local universe
where they could be studied in detail.  Thus there are limited
prospects for measuring the masses of specific elements in the ejecta.
An alternative approach relies on identifying signatures of elements
in a wide set of stars and galaxies.  Such observations try to
reconstruct individual nucleosynthesis elements from the imprint they
leave in stars formed immediately after the explosion took place.
Parallel efforts aim at putting constraints on the contributions
of the nucleosynthesis processes and sites to the enrichment of the
galaxy based on ratios of elements observed at different
metallicities, \ie, in environments formed in different cosmological
epochs.

The conditions that make jet-driven supernovae promising r-process
sites apply equally or even more so for the mergers of an NS and
another compact object, NS or BH.  Many theoretical and numerical
studies had been confirming this view already before it was
unequivocally demonstrated by the event observed in gravitational
waves and across the electromagnetic spectrum on 17/08/2017 and
thereafter \citep{Abbott_et_al__2017__PhysicalReviewLetters__GW170817ObservationofGravitationalWavesfromaBinaryNeutronStarInspiral,Abbott__2017__apjl__MultiMessengerObservationsofaBinaryNeutronStarMerger,Abbott__2017__TheAstrophysicalJournal__GravitationalWavesandGammaRaysfromaBinaryNeutronStarMergerGW170817andGRB170817A,Abbott__2017__TheAstrophysicalJournal__OntheProgenitorofBinaryNeutronStarMergerGW170817,Abbott__2017__TheAstrophysicalJournal__EstimatingtheContributionofDynamicalEjectaintheKilonovaAssociatedwithGW170817,Tanvir__2017__TheAstrophysicalJournal__TheEmergenceofaLanthanideRichKilonovaFollowingtheMergerofTwoNeutronStars}.  Since then, mergers have a solid observational foundation
as r-process sites that the competing model of rare CCSNe still
lacks.

\section{Dynamics of jets}
\label{Sek:Dyn}

\begin{figure}
  \centering
  \includegraphics[width=0.8\linewidth]{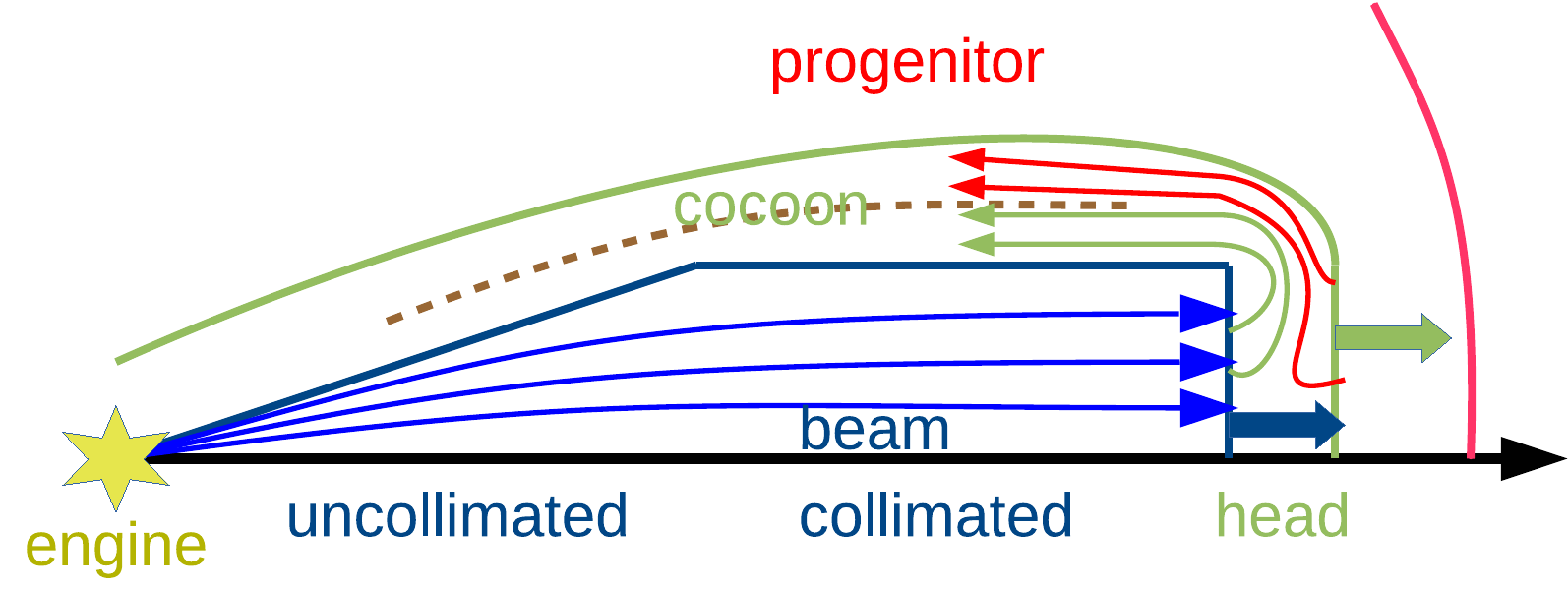}
  \caption{
    Simple representation of a jet in a star whose surface is indicated by the red curve.  Symmetry about the propagation direction (black axis) is assumed.  Gas is accelerated by a central engine (yellow symbol) and follows the velocity field indicated by the thin blue and green arrows as long as it is in the beam and the head/cocoon, respectively.  Additionally, progenitor matter entering the cocoon via the head is indicated by the red arrows. Expanding to the right (thick arrows), the jet has uncollimated (conical) and collimated (cylindrical) sections.
  }
  \label{Fig:jetstructure}
\end{figure}

Stars tend to be round.  Developing a jet and
thus blatantly violating this rule requires special conditions that
are not easily met even in violent events such as CCSNe.  Observations
as well as numerical simulations of ordinary CCSNe show deviations
from spherical geometry caused by the hydrodynamic instabilities
playing an important role in the explosion mechanism \citep[][]{Mueller__2020__LivingReviewsinComputationalAstrophysics__HydrodynamicsofCoreCollapseSupernovaeandTheirProgenitors}.  However, the
degree of asymmetry is lower than what can be inferred for many
more extreme, rare events from, \eg, the velocity distribution of the
ejecta, the polarization of the emitted radiation, or the connection
between CCSNe and relativistic GRBs
\citep[][]{Piran__2004__RMP__The_physics_of_GRBs,Woosley_Bloom__2006__araa__The_Supernova_Gamma-Ray_Burst_Connection,Piran__2019__TheAstrophysicalJournal__RelativisticJetsinCoreCollapseSupernovae,Corsi__2021__NewAstronomyReviews__GammaRayBurstJetsinSupernovae}.

These observations can be understood in terms of rapid outflows with a
narrow cylindrical or conical shape which can be described
approximately in axisymmetry as in the schematic picture shown in
\figref{Fig:jetstructure} \citep[see, e.g.,][]{Bromberg_et_al__2011__apj__ThePropagationofRelativisticJetsinExternalMedia}.  This structure is not stationary, but
keeps expanding radially at a rate that can be expressed in terms of the
pattern speed of the head of the jet.  Gas is injected into the jet at
its base and then propagates along the \emph{beam} towards the head of
the jet at high (supersonic or, for magnetized jets, super-fast)
speeds exceeding the pattern speed of the jet.  As a consequence, the
vertical motion of gas is stopped at the jet head and a terminal or
reverse shock forms. Passing through the shock, the gas heats up and
expands sideways. It forms the hot, tenuous inner \emph{cocoon} where
the gas velocity is directed inward, at least in the frame comoving
with the beam. As the jet propagates, it compresses its environment and
creates a second, bow shock ahead of the terminal shock.
The outer cocoon region between the bow shock and the inner cocoon is
filled by, compared to the inner cocoon, denser shocked gas from the
surrounding layers, with a contact discontinuity separating the two
components of the cocoon.

In the context of CCSNe, jets are launched at the
center of the collapsed core.  Hence, they will propagate into a dense
medium (in contrast to, \eg, jets in binary stars or active galactic
nuclei).  This medium can collimate the jet
\citep[][]{Matzner__2003__MonthlyNoticesoftheRoyalAstronomicalSociety__SupernovaHostsforGammaRayBurstJetsDynamicalConstraints,Lazzati_et_al__2012__apj__UnifyingtheZooofJet-drivenStellarExplosions,Bromberg_et_al__2011__apj__ThePropagationofRelativisticJetsinExternalMedia}.
In this scenario, the lateral jet structure depends on the (ram)
pressure balance between the jet, the cocoon, and the external medium,
while its propagation speed is affected by that between the ram
pressures exerted by the beam and the external medium onto each other
across the jet head.  The resulting interplay, which depends on the
profile of the external medium as well as the jet parameters,
determines whether the jet has a collimated, cylindrical or an
uncollimated, conical shape and also sets its propagation speed and
energy flux.
Depending on the ratio between magnetic, rotational, radial kinetic and internal energy, the structure of a jet launched from a rotating, magnetized base is set at least partially by the centrifugal force and the Lorentz force of the magnetic field \citep[see, e.g.,][]{Komissarov__2021__NewAstronomyReviews__NumericalSimulationsofJets}.  The latter can have two opposite consequences: its isotropic pressure exerts a decollimating force, while the hoop stress of the toroidal field component ($B_{\phi}$ in a cylindrical coordinate system whose $z$-axis is aligned with the jet) has a pinching effect on the gas.  
The magnetic field strength tends to decrease steeply with distance from the center such that, even if the base is strongly magnetized, the asymptotic regime of the jet far away from its source will not be dominated by the magnetic field. 

Due to the combination of the forward motion and the (differential)
rotation of the jet, the magnetic field lines injected with the gas
typically show a helical shape.  Despite the pinching force of the
toroidal component, the magnetic self-collimation of such a
configuration may not be effective under general conditions \citep{Lyubarsky__2009__TheAstrophysicalJournal__AsymptoticStructureofPoyntingDominatedJets}.  Acting
similarly to the elastic force in a rubber band, the magnetic tension
force with its tendency to shorten the magnetic field lines can
transmit the inertia of the fluid elements threaded by a field line \cite{Begelman__1995__ProceedingsoftheNationalAcademyofScience__TheAccelerationandCollimationofJets}.
If the magnetic field lines are anchored at the jet base, \ie, in an
accretion disk or the surface of a PNS, the connection of the gas in
the jet with their footpoints can provide an additional mechanism for
preventing sideways expansion of the jet and thus contribute to the
collimation.  The force responsible for this process is transmitted at
the speed of an \Alfven wave and thus can only operate where the magnetic field dominates the dynamics.
Beyond this location, collimation relies mostly on the external medium.

To reach the stellar surface, a jet generated near the core has to
traverse several orders of magnitude in radius and ambient density.
To achieve this, the jet power and velocity have to be sufficiently
high \cite{Aloy__2018__MonthlyNoticesoftheRoyalAstronomicalSociety__OntheExistenceofaLuminosityThresholdofGRBJetsinMassiveStars} and the central engine powering it has to be active
for a sufficiently long time \citep{Bromberg_et_al__2015__mnras__OnthecompositionofGRBsCollapsarjets}.  The conditions in the
jet may vary significantly during this period.  Variations at the base
of the jet translate into a possibly rich substructure along the jet,
\eg, shocks further heating the gas.  If the jet is too weak or the
engine active for too short a time, the jet may be quenched by the ram
pressure of the stellar matter and turn into a wide, slower
(sub-relativistic) outflow.

The typical velocity and magnetic fields of jets are prone to several
types of instabilities\citep[see, e.g.,][]{Komissarov__2021__NewAstronomyReviews__NumericalSimulationsofJets}.  Depending on the topology of the magnetic
field, current-driven instabilities may cause the growth of
non-axisymmetric kink modes, which can displace the jet beam from the
$z$-axis.  Such a displacement corresponds to a deformation of the jet or,
if it is too large, may cause a complete disruption, thus preventing
the jet from breaking out of the star.

Neutrino heating is a key process for launching a regular CCSN \citep[][]{Woosley__1993__TheAstrophysicalJournal__GammaRayBurstsfromStellarMassAccretionDisksaroundBlackHoles,MacFadyen_Woosley__1999__ApJ__Collapsar,Nagataki_et_al__2007__apj__GRB_Jet_Formation_in_Collapsars}.  The neutrinos are emitted by the PNS and the gas
accreted by it.  The emission, while modified by the stochastic nature
of the flows, is roughly spherical.  On the other hand, the centrifugal force can transform an
extremely rapidly rotating core into an oblate configuration, ranging
from a rotationally flattened PNS to the development of a
centrifugally supported disk surrounding the central compact object.
The anisotropy of the system and, in particular, of its
neutrinospheres enhances the emission in the direction of the
rotational axis and suppresses it in the equatorial plane.  Neutrino
heating is thus strongest near the rotational axis and may therefore drive a
polar outflow.  

In the collapsar model, a low-density funnel develops along the rotational axis near the newly formed BH.  Neutrino-antineutrino pairs emitted by the accretion disk can annihilate in this funnel and provide the thermal energy required to launch jets.  
In contrast to the neutrino reactions that are crucial for regular CCSNe, pair processes can heat the gas without changing its electron fraction.
Hence, if gas from a very neutron-rich environment such as the PNS or the disk ends up in the jet, it may retain its initial low electron fraction, possibly enabling r-process nucleosynthesis.

If the magnetic field dominates the dynamics, it will force the
fluid to move along the field lines.  Hence, a field with a
large-scale component aligned with the $z$-axis may direct gas in that
direction and thus favor the formation of a jet.  
{Gradients of the magnetic pressure along the field lines drive gas from regions of strong field to weaker magnetized layers and provide the acceleration of the outflow.}
{Such a configuration can be found in a core threaded by a uniform magnetic field aligned with the rotational axis.  The high densities of the PNS or accretion disk force the magnetic field lines to follow the fluid flow, whose}  
differential rotation twists the magnetic field lines
into a helical shape, starting near the PNS or
disk and gradually extending upward.  The whole process generates a
toroidal field component, $B^{\phi}$, from the vertical one and
increases the field strength and magnetic pressure.  The pressure
gradient in turn accelerates gas away from the center while the
magnetic tension restricts motion across the field lines.  Versions of
this \emph{magnetic tower} geometry have been considered for
astrophysical systems from protostellar objects to active galactic
nuclei as well as stellar core collapse 
\citep{Lynden-Bell__1996__mnras__Magneticcollimationbyaccretiondiscsofquasarsandstars,Lynden-Bell__2003__mnras__Onwhydiscsgeneratemagnetictowersandcollimatejets,Uzdensky__2006__apj__Stellar_Explosions_by_Magnetic_Towers,Uzdensky_MacFadyen__2007__apj__Magnetar-Driven_Magnetic_Tower_as_a_Model_for_Gamma-Ray_Bursts_and_Asymmetric_Supernovae}.

This process illustrates the ability of the magnetic field to transport angular
momentum along field lines, in this case from the rapidly rotating PNS
or disk to the slower outer layers.  
Rotational energy is extracted from the central object and converted in a slingshot effect into
(radial/vertical) kinetic energy of the outflow.
{In a differentially rotating disk that is connected to the outer layers of the star by a magnetic field, gas} just above the disk surface can be flung outward by the centrifugal force along a field
line.
{The geometry of this outflow is that of an equatorial wind, but it could also be redirected along the axis by the collimation mechanisms mentioned above \citep{Metzger_et_al__2007__apj__Proto-NeutronStarWindswithMagneticFieldsandRotation,Bucciantini_et_al__2007__mnras__Magnetar-driven_bubbles_and_the_origin_of_collimated_outflows_in_GRBs}.}

Both basic magnetic acceleration mechanisms rely on a large-scale
ordered magnetic field.  Its presence, however, cannot be taken for granted.  Even if
the progenitor possesses an ordered field, \eg, of dipole geometry, it
may be subdominant \wrt the small-scale turbulent field generated by
instabilities such as convection or the magneto-rotational
instability.  Still, a large-scale dynamo may operate and generate a
strong large-scale field if the turbulent system is rotating very
rapidly, thus allowing for jet formation via the aforementioned
mechanisms \citep[e.g.,][]{Duncan_Thompson__1992__ApJL__Magnetars,Thompson_Duncan__1993__ApJ__NS-dynamo,Moesta_et_al__2015__nat__Alarge-scaledynamoandmagnetoturbulenceinrapidlyrotatingcore-collapsesupernovae,Raynaud__2020__ScienceAdvances__MagnetarFormationthroughaConvectiveDynamoinProtoneutronStars,ReboulSalze__2021__aap__AGlobalModeloftheMagnetorotationalInstabilityinProtoneutronStars,Lander__2021__mnras__GeneratingNeutronStarMagneticFieldsThreeDynamoPhases,White__2022__TheAstrophysicalJournal__OntheOriginofPulsarandMagnetarMagneticFields}.

Matter in a disk spirals onto the central object due to an outward
transport of angular momentum, leading to a loss of centrifugal
support.  This transport can be described in terms of an effective
viscosity \cite{Shakura_Sunyaev__1973__AA__alpha_visco} as a
phenomenological model for small-scale turbulence.  The main driver of
this turbulence is the magnetorotational instability (MRI)
\cite{Balbus_Hawley__1998__RMP__MRI} to which (Keplerian) accretion
disks are unstable.  Apart from amplifying a weak seed field to a
relevant strength, the MRI may also form an ingredient of a
large-scale dynamo.  

Finally, rotational energy extracted from the BH via the Blandford-Znajek process
\cite{Blandford_Znajek__1977__mnras__Electromagnetic_extraction_of_energy_from_Kerr_black_holes},
an MHD version of the Penrose process, can power jets.  To operate
this process, the ergosphere of the rotating BH has to possess a low
gas density.  Under these conditions, it can, for parameters
compatible with those in post-collapse stellar cores, sustain a jet
power consistent with the requirements for GRBs \cite{Komissarov_Barkov__2009__mnras__ActivationoftheBlandford-Znajekmechanismincollapsingstars}.

\section{Nucleosynthesis processes}

The goal of a nucleosynthesis model is to describe the (final)
composition of the exploding star, \ie, the abundances $Y_{N}$ of a large set of nuclei $\{N\}$.  These change due to nuclear reactions
involving the nucleus $N$ itself and potential other reactants $J$
(other nuclei, but also photons and leptons)with which it reacts.
The reaction cross sections depend on the energy of the
individual particles.  If the particles involved in a reaction are in
local thermodynamic equilibrium, they obey the
respective equilibrium distribution \citep[\eg, Planck or
Maxwell-Boltzmann][]{Cox1968,Clayton1968,Iliadis2007,Lippuner2017,Arnould2020,Cowan__2021__ReviewsofModernPhysics__OriginoftheHeaviestElementstheRapidNeutronCaptureProcess}.  
Consequently, reactions between
nuclei, electrons, and photons depend on density and gas temperature.
For less dense regions that are located at larger distances to the central object, neutrinos
are not tightly coupled to the gas.  The rates of neutrino emission
and absorption depend on the neutrino spectra, which have to be known
explicitly or parametrized by, \eg, neutrino temperatures which may
differ from that of the gas
\citep{Tamborra2012,Lippuner2017,Sieverding2019}.  If all these
variables are known, a given initial condition will evolve according
to a large system of coupled ordinary differential equations, a so-called nuclear reaction network.

Uncertainties in the cross sections due to nuclear physics have
to be taken into account. The cross sections are experimentally
better accessible close to the valley of stability in the nuclear
chart and, as a tendency, reactions involving nuclei far from it are less well known. In addition, extrapolations of experimentally
determined reaction rates to the relevant temperatures can impose uncertainties \citep[see, e.g.,][for a review]{Horowitz2019}.

Noticeable, direct modifications of the cross sections in the
presence of a strong magnetic field is neglected in reaction
networks. Indirectly, however, the field influences the
nucleosynthesis by changing the dynamics and consequently the density and temperature. However, for extreme conditions with magnetic field strength of $\gtrsim 10^{15}\,\mathrm{G}$, the magnetic field can also directly influence the binding energies of the nuclei and therefore the composition \citep{Kondratyev2018,Kondratyev2019}. The necessary
large magnetic field strength may be found in the center of a strongly magnetized CC-SNe \citep[e.g.,][]{Moesta_et_al__2015__nat__Alarge-scaledynamoandmagnetoturbulenceinrapidlyrotatingcore-collapsesupernovae,Obergaulinger_Aloy__2017__mnras__Protomagnetarandblackholeformationinhigh-massstars}, accompanied with high temperatures and densities.

If the reactions occur on much shorter times than the thermodynamic
conditions changes due to other factors such as the dynamics of the
system, the nuclei are in nuclear statistical equilibrium (NSE).  In
this case, the abundances can be obtained by solving the Saha equation
together with the constraints of mass and charge conservation
\citep{Clifford1965,Hartmann1985,Hix__1999__JournalofComputationalandAppliedMathematics__ComputationalMethodsforNucleosynthesisandNuclearEnergyGeneration.,Iliadis2007,Seitenzahl2008,Lippuner2017}.
The NSE composition is fully determined by nuclear properties, temperature, density, and electron fraction, i.e., all
memory of an earlier composition is lost. In stellar explosion, this is the case for high temperatures of $T \gtrsim 6 \, \mathrm{GK}$ that regularly occurs in the PNS and its surroundings and behind the
CCSN shock wave.

This effect causes a dichotomy in the ejected gas.  If a fluid
element at some point reaches NSE temperatures, its pre-collapse
composition will be erased and replaced by NSE. Later on, it may cool down enough to
leave the NSE regime. From this point on, the non-equilibrium
reaction rates will determine the further evolution starting from the last NSE composition. The final abundances of fluid elements whose maximum temperature remains below the NSE threshold, on the other hand, depend on the progenitor composition and hence on stellar evolution models.

The abundances strongly depend on the peak temperature and density as well as the electron fraction and entropy of the gas. The observationally very important
$^{56}$Ni, e.g., requires temperatures $T_\mathrm{peak}\gtrsim4\,\mathrm{GK}$, which
can be found at the shock front or in matter ejected from close to the
central PNS.  For a higher explosion energy, the shock wave will be
able to maintain a higher temperature farther outside and the amount
of ejected matter with $T_\mathrm{peak}\ge4\,\mathrm{GK}$ is therefore
larger \citep{Woosley_Heger_Weaver__2002__ReviewsofModernPhysics__The_evolution_and_explosion_of_massive_stars}, causing the $^{56}$Ni mass to correlate
with the explosion energy
\citep[e.g.,][]{Maeda2003,Nomoto_et_al__2013__araa__NucleosynthesisinStarsandtheChemicalEnrichmentofGalaxies,Chen2017,Nomoto2017,Suwa2019,Grimmett__2021__MonthlyNoticesoftheRoyalAstronomicalSociety__TheChemicalSignatureofJetDrivenHypernovae}.
In addition to the peak temperature, the neutron-richness influences the amount of synthesized $^{56}$Ni. More neutron-rich conditions
lower the amount of synthesized $^{56}$Ni
\citep{Nishimura_et_al__2015__apj__Ther-processNucleosynthesisintheVariousJet-likeExplosionsofMagnetorotationalCore-collapseSupernovae,Chen2017,Grimmett__2021__MonthlyNoticesoftheRoyalAstronomicalSociety__TheChemicalSignatureofJetDrivenHypernovae}. In fact, the dominant
mass of $^{56}$Ni is synthesized under symmetric or proton-rich
condition ($Y_e \ge 0.5$).

The heaviest elements can be formed by successive captures of
neutrons on seed nuclei.  A sequence of several such reactions leads
the nucleus away from the valley of stability in the nuclear chart.
Once $\beta$-decays are faster than neutron captures, the nucleus will
start decaying back to a more symmetric neutron-to-proton ratio.  Each of these excursions increases the mass number.  For sufficiently high neutron fluxes, the resulting rapid neutron-capture process or
r-process for short (to distinguish it from the s-process with slower
neutron captures that can take place in evolved post-main-sequence
stars) can synthesize elements up to the actinides. The magic numbers
of closed neutron shells at $N = 50, 82, 126$ can translate to three peaks of enhanced abundances, the so called first, second, and third r-process peaks, respectively.

The detailed final abundance pattern, in particular whether a
system does indeed reach all three of these potential peaks
or only the first one or two, depends on the precise conditions.  The
higher the ratio between neutrons and seed nuclei is, the stronger the
r-process will be \citep[e.g.,][]{Freiburghaus1999}.  Under many conditions, the r-process takes place
in matter that was previously in NSE, for which conditions of high
entropy or low $Y_e$ enhance the fraction of free neutrons.

The neutron-richness is set by positron- and electron- as well as
neutrino reactions:
\begin{align} 
  e^- + p &\rightleftarrows n + \nu_e, \\
  e^+ + n &\rightleftarrows p + \bar{\nu}_e,
\end{align} 
where reactions on nucleons are considered to dominate
over reactions on heavier nuclei. The evolution of the neutron-richness at high temperatures can therefore be expressed as
\begin{equation} 
  \frac{\mathrm{d} Y_e}{\mathrm{d} t} =
  (\lambda_{e^+}+\lambda_{\nu_e}) Y_n -
  (\lambda_{e^-}+\lambda_{\bar\nu_e})Y_p \, ,
\end{equation} 
where $\lambda_{S}$ stands for the rate of captures of particle
species $S$.  

In the densest environments, some or all of these reactions can be
sufficiently fast for an equilibrium between neutrinos and matter
which sets the value of electron fraction depending on the
thermodynamic conditions and the fluxes of (anti-)neutrinos \citep[for
a thorough discussion,
see, e.g.,][]{Just__2022__MonthlyNoticesoftheRoyalAstronomicalSociety__NeutrinoAbsorptionandOtherPhysicsDependenciesinNeutrinoCooledBlackHoleAccretionDiscs}.
After the initial $\nu_e$-burst, $\nu_e$ and $\bar{\nu}_e$ are emitted
by the PNS with very similar luminosities and only slightly different energies. The equilibrium electron fraction set by neutrino reactions only (i.e., neglecting electron and positron captures) is hereby
close to $Y_e \approx 0.5$ \citep[e.g.,][]{Qian__1996__TheAstrophysicalJournal__NucleosynthesisinNeutrinoDrivenWinds.I.thePhysicalConditions,Arcones2013,Just__2022__MonthlyNoticesoftheRoyalAstronomicalSociety__NeutrinoAbsorptionandOtherPhysicsDependenciesinNeutrinoCooledBlackHoleAccretionDiscs}. Initially neutron rich matter
exposed to these conditions will thus approach more proton-rich conditions.  For
matter to be unbound from the potential well of the PNS solely by
neutrino heating, many reactions with neutrinos are required. Even if the aforemention values of $Y_e \approx 0.5$ are never reached, the resulting conditions most likely are not neutron-rich enough to host a succesful r-process, thus making regular CCSNe unlikely to eject heavy elements.

A lower electron fraction will be obtained for matter that got ejected with as few neutrino reactions involved as possible, e.g., due to an additional magnetic pressure contribution. The ratio between heating by
neutrinos and energy input by the magnetic field will therefore set
the neutron-richness of the outflowing matter. 

Another process to synthesize heavier elements, the so-called n-process, can be triggered when the shock passes through the He layer of the star \citep{Blake1976,Truran1978,Thielemann1979,Blake1981,Hillebrandt1981,Rauscher2002}. There, it can induce the reaction $\AZNucleus{22}{}{Ne}(\alpha,n)\AZNucleus{25}{}{Mg}$ and neutrons are released. The resulting neutron burst will modify the
pre-explosion abundance pattern and possibly shift them to higher mass numbers. The strength of the neutron burst depends on the peak temperature in the He layer and therefore on the explosion energy.
\cite{Choplin__2020__aap__AStrongNeutronBurstinJetlikeSupernovaeofSpinstars}
considered the case in which rotational mixing during the stellar
evolution has boosted the enrichment of s-process nuclei in the outer
layers of the star. They studied a highly energetic, relativistic jet
injected into such a layer and found that the neutronization is stronger
than in the case of an equivalent spherical explosion.  While some of the
heavy s-process nuclei may act as neutron poison, thus preventing an
r-process, they are nevertheless shifted to higher masses since the temperatures are low enough to suppress photodisintegration.

While the aforementioned processes are responsible for heavy and neutron-rich nuclei, it should be noted that other processes exist that can synthesize proton-rich isotopes as found within our Sun \citep[e.g.,][]{Cameron1957,Meyer2000,Rauscher2013,Pignatari2015,Lugaro2016,Bliss2018,Eichler2018}. One of them is based on captures of protons \citep[p-process, e.g.,][]{Burbidge1957,Audouze1975,Meyer1994,Arnould2003}. The captures of protons has, however, been found to require special and possibly not available conditions. Therefore, the understanding of the p-process has changed over time. Nowadays the p-process (or also often called $\gamma$-process) is not understood as captures of protons, but describes the photodissociation of pre-existing heavier seed nuclei under moderately hot conditions when the shock moves through the layers of the exploding star \citep[e.g.,][]{Arnould1976,Woosley1978,Rayet1995,Pignatari2016,Choplin__2020__aap__AStrongNeutronBurstinJetlikeSupernovaeofSpinstars}. Neutrinos can play an additional role in synthesizing moderately heavy proton-rich nuclei via the $\nu$p-process \citep{Froehlich2006,Pruet2006,Wanajo2006}. Within this process antineutrino absorption on protons produce neutrons that are captured immediately afterwards on heavier (proton-rich) nuclei, therefore bypassing bottleneck reactions.

\section{Numerical methods}
\label{Sek:Num}

\subsection{Dynamical simulations}
\label{sSek:nuMHD}

Key properties such as the ejecta--mass, energy, velocity, thermodynamics, and morphology depend on the interplay of the dynamics of the gas and the magnetic field, relativistic gravity, nuclear physics, and the emission as well as the transport of neutrinos.  Including all of these ingredients in a numerical model
is a complex task that converts CCSN simulations into some of the
computationally most expensive projects realized on large
supercomputers.  Research on regular CCSNe has seen tremendous
progress as various groups managed to perform simulations with a full
set of input physics in three spatial dimensions, \ie, without the
previously necessary assumptions of spherical or axial symmetry
\citep[\eg,][]{Lentz2015,Janka_et_al__2016__AnnualReviewofNuclearandParticleScience__PhysicsofCore-CollapseSupernovaeinThreeDimensionsASneakPreview,Roberts2016,Takiwaki2016,Oconnor2018,Mueller__2020__LivingReviewsinComputationalAstrophysics__HydrodynamicsofCoreCollapseSupernovaeandTheirProgenitors,Burrows__2020__MonthlyNoticesoftheRoyalAstronomicalSociety__TheOverarchingFrameworkofCoreCollapseSupernovaExplosionsAsRevealedby3DFORNAXSimulations}.
After a long time during which their physical sophistication had been
lagging behind those of regular ones, simulations of more exotic,
potentially jet-forming CCSNe have recently become more reliable \citep[e.g.,][]{Winteler_et_al__2012__apjl__MagnetorotationallyDrivenSupernovaeastheOriginofEarlyGalaxyr-processElements,Moesta_et_al__2015__nat__Alarge-scaledynamoandmagnetoturbulenceinrapidlyrotatingcore-collapsesupernovae,Kuroda_et_al__2020__apj__MagnetorotationalExplosionofaMassiveStarSupportedbyNeutrinoHeatinginGeneralRelativisticThreeDimensionalSimulations,Obergaulinger__2021__mnras__MagnetorotationalCoreCollapseofPossibleGRBProgenitorsIII.ThreeDimensionalModels}.  

Simulations of CCSNe are usually initialized using the results of
one-dimensional, hydrostatic stellar evolution models at the onset of
collapse.  Some of these calculations include approximate
prescriptions of the effects of rotation and magnetic fields such
as the Tayler-Spruit dynamo
\citep[\eg,][]{Spruit__2002__AA__Dynamo,Heger_et_al__2005__apj__Presupernova_Evolution_of_Differentially_Rotating_Massive_Stars_Including_Magnetic_Fields}.
However, even in that case the radial profiles of these important
input variables and their angular distribution are highly uncertain.
Besides this problem of how to map spherical progenitor data to the
multidimensional grid of the CCSN simulation, a major limitation of
this approach is the suppression convective instabilities that would
otherwise generate large-scale deformations in layers to be accreted
onto the PNS or BH.  While these deformations can facilitate
neutrino-driven explosions
\citep{Couch_Ott__2013__apjl__RevivaloftheStalledCore-collapseSupernovaShockTriggeredbyPrecollapseAsphericityintheProgenitorStar,Varma__2021__MonthlyNoticesoftheRoyalAstronomicalSociety__3DSimulationsofOxygenShellBurningwithandwithoutMagneticFields,Bollig__2021__TheAstrophysicalJournal__SelfConsistent3DSupernovaModelsfrom7Minutesto7Sa1BetheExplosionofa19MsolProgenitor},
their impact on the formation of jets, \eg, via convective
amplification of the magnetic field, has not been investigated.

The codes used for evolving these initial conditions through collapse
and the subsequent phases combine some or all of the following
ingredients:
\begin{itemize}
\item The evolution of the gas and the magnetic field is described by
  relativistic or Newtonian (M)HD, usually on two- or
  three-dimensional grids in spherical or Cartesian coordinates.  One
  of the major unsolved challenges lies in resolving processes that
  occur at small length scales, \eg, the turbulent amplification of
  the magnetic field.
\item Ideally, a fully general relativistic (GR) treatment of gravity
  is used.  For the sake of simplicity, however, many codes are based
  on pseudo-relativistic approximations and employ corrections to the
  Newtonian gravitational potential.  These potentials are calibrated
  such as to reproduce crucial properties of PNSs or BHs computed in
  full GR.
\item All recent simulations describe the matter at high densities by
  a nuclear EOS available in tabulated form.  The table may be based
  on various methods for describing the nuclear structure and
  interactions, some including effects such as the formation of
  hyperons.  They assume nuclear statistical equilibrium (NSE) to obtain
  the composition of the baryonic matter as a function of density,
  temperature, and electron fraction.  At lower densities and
  temperatures, where NSE breaks down, an EOS accounting
  for leptons, photons, and nuclei can be used.  The best option for
  obtaining the chemical composition of the gas is to follow the
  transport of and the reactions between all nuclei making up the
  stellar matter.  This ideal, based on nuclear reaction networks (see
  below), is too expensive to be put in practice.  Common ways around
  this difficulty are to apply an approximate, ad-hoc composition,
  \eg, by applying NSE also in this regime, or to solve the evolution
  of only a reduced subset of nuclei.
\item An accurate treatment of neutrinos accounts for the largest
  fraction of the computational time consumed, mostly because the
  neutrino field should be described in seven-dimensional phase space
  of four space-time and three momentum dimensions.  Since so far
  such a scheme remains an elusive goal, all codes used for CCSNe
  employ simplified methods ranging from leakage schemes that avoid
  transport altogether to ones based on the transport equations for
  the lowest one or two angular moments of the neutrino phase-space
  distribution, so-called $M0$ or $M1$ schemes, respectively. This allows for a much more reliable determination of the
  entropy and electron fraction than more approximate methods.  It
  should be noted that, when it comes to modeling jets generated by
  neutrino heating, even $M1$ methods meet their limitations as they
  may produce large errors in the annihilation rates of intersecting
  rays of neutrinos and antineutrinos.
\item A wide set of reactions between neutrinos and matter has to be
  included to get the dynamics as well as the composition of the
  ejecta right such as (inverse) $\beta$-processes, scattering off
  baryons and leptons, and pair processes.  
\end{itemize}

Numerical simulations at the level of detail outlined above can be run for at most a few seconds, which is far too short to follow 
{how the ejecta propagate to the stellar surface and into the circumstellar medium as well as how hydrodynamic instabilities mix the chemical elements during the propagation}.  
Specific techniques have been designed to evolve the ejecta for many hours, days, or even months by gradually switching to more and more approximate descriptions of the most expensive physics.  While these methods have been used successfully to compare models for regular CCSNe to observations of, \eg, SN 1987A or the Cas A nebula
\citep[\eg][]{Wongwathanarat_et_al__2015__aap__Three-dimensionalsimulationsofcore-collapsesupernovae:fromshockrevivaltoshockbreakout,Alp__2019__TheAstrophysicalJournal__XRayandGammaRayEmissionfromCoreCollapseSupernovaeComparisonofThreeDimensionalNeutrinoDrivenExplosionswithSN1987A,Stockinger__2020__MonthlyNoticesoftheRoyalAstronomicalSociety__ThreeDimensionalModelsofCoreCollapseSupernovaefromLowMassProgenitorswithImplicationsforCrab,Orlando__2021__aap__TheFullyDevelopedRemnantofaNeutrinoDrivenSupernova.EvolutionofEjectaStructureandAsymmetriesinSNRCassiopeiaa,Gabler__2021__MonthlyNoticesoftheRoyalAstronomicalSociety__TheInfancyofCoreCollapseSupernovaRemnants},
similarly self-consistent studies for jet-driven CCSNe are not yet as
far developed.

More effort has gone into simulating the propagation
of artificially launched jets through the star.  A large body of
research exists based on relativistic (M)HD simulations (mostly)
with simplified neutrinos and nuclear physics in which the central compact
object has been excised and replaced by an inner boundary condition
through which a jet with prescribed parameters (density, entropy,
velocity, potentially magnetization) is injected.  While they cannot
be used to study the physics of the formation of the jet, such models
are useful to assess the jet propagation and stability, the mixing of
elements, and the electromagnetic signal at jet breakout.

\subsection{Nuclear reaction networks and tracer particles}

Following the nuclear reactions amounts to solving the equations of
the nuclear network coupling the evolution of all isotopes that are of
interest.  Mathematically, such a network is a set of coupled ordinary
differential equations in which each reaction between $n = 1, 2, 3..,$
nuclei is represented by a coupling term proportional to the
abundances of the $n$ nuclei involved and the reaction rates.  Since
the time scales corresponding to the reactions can span orders of magnitudes, the
system is solved commonly using implicit time stepping methods
\citep[e.g.,][]{Woosley1973,Arnould1976,Thielemann1979,Hix__1999__JournalofComputationalandAppliedMathematics__ComputationalMethodsforNucleosynthesisandNuclearEnergyGeneration.,Timmes1999,Hix2006,Kostka2014,Longland2014,Lippuner2017}.
The computational costs of a network scale non-linearly with the
number of isotopes. Regimes in which NSE holds are an exception to this rule since in this case the abundances can be obtained by a set of algebraic Saha equations, thus reducing the computational effort.

For determining the spatio-temporal distribution of the nuclear
species, an \emph{in-situ} network can be coupled to a (M)HD simulation code.
In this case, additional terms describing the advection of the
isotopes with the fluid flow are included that turn the system into one of partial differential equations.  The advantages of such a strategy are that the thermodynamic conditions and the neutrino fields affecting the reaction rates can be directly taken from the simulation and that the thermal energy liberated or consumed by the reactions can directly feed back onto the dynamics \citep[e.g.,][]{Mueller1986,Cabezon2004,Bruenn2006,Nakamura2014}. However, the computational costs of solving the dynamics together with the network can be prohibitive. Therefore, only a reduced subset of species that are
responsible for most of the nuclear energy output are included.
Common reduced networks use around 20, only occasionally up to more than 100,
species \citep[e.g.,][]{Weaver1978,Timmes1999,Paxton2011,Harris2017,Sandoval2021}.

Including all species, some of which are of great interest
observationally, is not
feasible, in particular when one takes into account that nuclear
reactions can go on for much longer time than dynamical times of an
explosion.  Full networks with up to several 1000s of isotopes are
usually run in post-processing after the dynamical simulation has been
completed.  This approach takes advantages of the fact that most
reactions have a negligible effect on the dynamics because they
contribute very little to the energy budget of the fluid.  Thus, if
it is known how a fluid element evolves through real space and through the
parameter space of density, temperature, electron fraction, neutrino
fluxes, and possibly a few other variables, the nucleosynthesis can be
computed accurately.  What is needed are thus Lagrangian tracer particles (often also called \emph{tracers} or \emph{trajectories}) of fluid
elements representing a given mass of the ejected gas.  Simplified
studies parameterize these trajectories using fits to numerical
simulations or analytic results for the expansion of the ejecta.
In contrast to simple parameterized estimates of the thermodynamic conditions, (M)HD simulations can provide the prevailing conditions directly.

Simulation based on particles inherently trace their conditions in a Lagrangian way. However, most of the (M)HD simulations are performed on an Eulerian grid. In this case, the tracer particles can be initially placed on the grid using different placement criteria \citep{Bovard2017}. Within the simulation, the tracers passively follow the fluid flow by integrating the fluid velocities \citep[e.g.,][]{Nagataki1997,Seitenzahl2010,Nishimura_et_al__2015__apj__Ther-processNucleosynthesisintheVariousJet-likeExplosionsofMagnetorotationalCore-collapseSupernovae,Harris2017,Bovard2017}. All particles interpolate and track the hydrodynamic quantities as well as the neutrino properties from the underlying grid. For each of the tracers, a nucleosynthetic evolution can be calculated separately in a so-called \emph{single-zone} approach. The assumption that the particles do not interact with each other is valid if the burning time scale is shorter than the diffusion time scale, which is the case for most explosive scenarios \citep{Hix__1999__JournalofComputationalandAppliedMathematics__ComputationalMethodsforNucleosynthesisandNuclearEnergyGeneration.}.
How many tracers are needed to accurately resolve the simulation has been investigated mainly in the context of type Ia SNe \citep{Travaglio2004,Brown2005,Seitenzahl2010,Townsley2016}. \citet{Seitenzahl2010} found that $128$ particles per dimension (i.e., 128 in 1D, $128^{2}$ in 2D, and $128^{3}$ in 3D) are necessary to obtain nucleosynthetic yields that are converged within 2\% of the mass fractions. This corresponds to an average tracer mass of around $\sim 10^{-4}\,\mathrm{M}_\odot$. This agrees with the finding of \citet{Nishimura_et_al__2015__apj__Ther-processNucleosynthesisintheVariousJet-likeExplosionsofMagnetorotationalCore-collapseSupernovae} in the context of MR-SNe. There, a total of $10000-50000$ tracers, i.e., $\sim 100-220$ tracer per dimension were necessary to obtain a converged result, which corresponds on average to $\sim4\cdot 10^{-5}-2\cdot 10^{-4}\,\mathrm{M}_\odot$ per tracer. 

In addition to the convergence of the calculations, other sources of uncertainty can arise in the post-processing of the nucleosynthesis yields. Thermodynamic conditions are usually required for longer than the simulation lasts \citep[e.g.,][]{Fujimoto__2008__TheAstrophysicalJournal__NucleosynthesisinMagneticallyDrivenJetsfromCollapsars,Korobkin2012,Harris2017,Eichler2018}. Thus, depending on the environment, some kind of extrapolation is employed. However, these parametric expansions do not capture future hydrodynamic variations and an uncertainty remains. On top of that, at the end of the simulation, it is not completely known which part of the mass is ultimately ejected and which is not. Commonly, a tracer is assumed to be ejected if its total energy is positive. Often other criteria are introduced for technical reasons, as it is not possible to extrapolate the thermodynamic conditions in the case of a negative radial velocity and an increasing temperature and density \citep{Harris2017}.

\section{Numerical results}
\label{Sek:Results}

\subsection{Supernova jets}
\label{sSek:SNJets}

\begin{figure}
 \includegraphics[width=0.95\columnwidth]{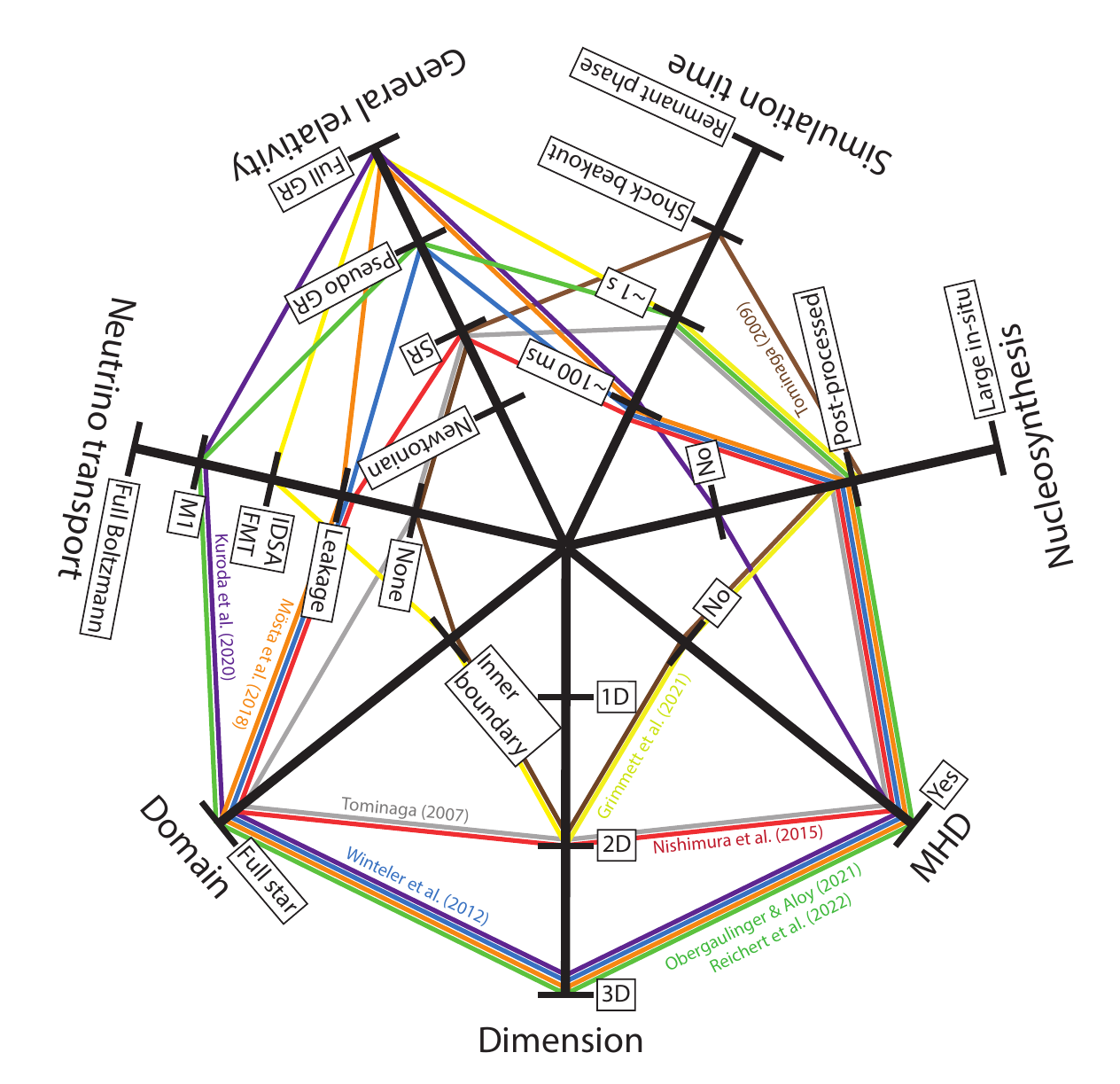}\\
 \caption{Overview of chosen physical inputs in different studies. Going inwards on the axis indicates a more approximate approach. Shown are the works of \citet[][grey]{Tominaga__2007__TheAstrophysicalJournal__TheConnectionbetweenGammaRayBurstsandExtremelyMetalPoorStarsBlackHoleFormingSupernovaewithRelativisticJets}, \citet[][brown]{Tominaga__2009__apj__Aspherical_Properties_of_Hydrodynamics_and_Nucleosynthesis_in_Jet-Induced_Supernovae}, \citet[][blue]{Winteler_et_al__2012__apjl__MagnetorotationallyDrivenSupernovaeastheOriginofEarlyGalaxyr-processElements}, \citet[][red]{Nishimura_et_al__2015__apj__Ther-processNucleosynthesisintheVariousJet-likeExplosionsofMagnetorotationalCore-collapseSupernovae}, \citet[][orange]{Moesta_et_al__2018__apj__r-processNucleosynthesisfromThree-dimensionalMagnetorotationalCore-collapseSupernovae}, \citet[][purple]{Kuroda_et_al__2020__apj__MagnetorotationalExplosionofaMassiveStarSupportedbyNeutrinoHeatinginGeneralRelativisticThreeDimensionalSimulations}, \citet[][yellow]{Grimmett__2021__MonthlyNoticesoftheRoyalAstronomicalSociety__TheChemicalSignatureofJetDrivenHypernovae}, and \citet[][green]{Obergaulinger__2021__mnras__MagnetorotationalCoreCollapseofPossibleGRBProgenitorsIII.ThreeDimensionalModels,Reichert2022}.}
 \label{fig:overview_studies}
\end{figure}

\begin{figure}
    \centering
    \includegraphics[width=\columnwidth]{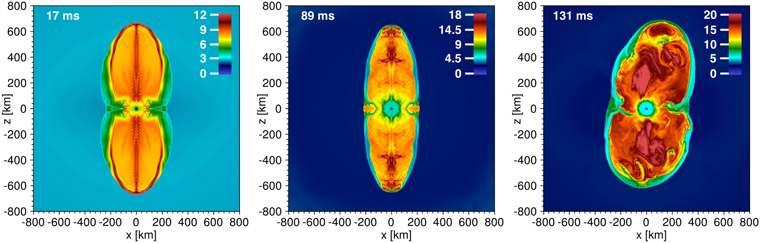}\\
    \includegraphics[width=\columnwidth]{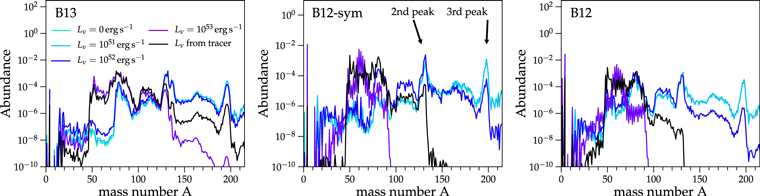}\\
    \caption{%
        Top panels: geometry of the specific entropy in three-dimensional models from \cite{Moesta_et_al__2018__apj__r-processNucleosynthesisfromThree-dimensionalMagnetorotationalCore-collapseSupernovae} after core bounce as indicated in the panels.  
        The color scale shows the specific entropy in meridional slices.  
        The models are, from left to right, a three-dimensional one with an initial magnetic field of $B_0 = 10^{13} \, \mathrm{G}$ (B13), and two simulations with a tenth of that initial field strength in octant symmetry (B12-sym) and unrestricted three-dimensional geometry (B12).
        Bottom panels: nucleosynthesis results for the same models.  Different line colors show results for tracer particles calculated assuming different neutrino luminosities.
        Figure taken with permission of the authors and the publisher from \cite{Moesta_et_al__2018__apj__r-processNucleosynthesisfromThree-dimensionalMagnetorotationalCore-collapseSupernovae} (\copyright~ IOP Publishing).
    }
    \label{fig:Moesta18Abund}
\end{figure}

Supernova progenitors do not directly collapse to BHs. Instead, all of them form a PNS onto which matter of relatively high density is
falling. With the dynamics of the core depending on processes inside
as well as around the PNS and on the interplay of matter, magnetic
field, and neutrino, this phase is most appropriately simulated
including the entire core and using full neutrino physics rather than,
e.g., excising the innermost region or simplifying the treatment of
neutrinos.  Simulations focusing on this stage of CCSNe have been
performed by, e.g.,
\cite{Burrows_etal__2007__ApJ__MHD-SN,Takiwaki_Kotake_Sato__2009__apj__Special_Relativistic_Simulations_of_Magnetically_Dominated_Jets_in_Collapsing_Massive_Stars,Scheidegger_et_al__2010__aap__Influence_of_model_parameters_on_GW_from_core_collapse,Takiwaki_Kotake__2011__apj__GravitationalWaveSignaturesofMagnetohydrodynamicallyDrivenCore-collapseSupernovaExplosions,Winteler_et_al__2012__apjl__MagnetorotationallyDrivenSupernovaeastheOriginofEarlyGalaxyr-processElements,Mosta_et_al__2014__apjl__MagnetorotationalCore-collapseSupernovaeinThreeDimensions,Nakamura__2015__aap__RProcessNucleosynthesisintheMHDneutrinoHeatedCollapsarJet,Tsujimoto_Nishimura__2015__apjl__TheRProcessinMagnetorotationalSupernovae,Nishimura_et_al__2015__apj__Ther-processNucleosynthesisintheVariousJet-likeExplosionsofMagnetorotationalCore-collapseSupernovae,Nishimura_et_al__2017__apjl__TheIntermediateRProcessinCoreCollapseSupernovaeDrivenbytheMagnetoRotationalInstability,Obergaulinger_Aloy__2017__mnras__Protomagnetarandblackholeformationinhigh-massstars,Moesta_et_al__2015__nat__Alarge-scaledynamoandmagnetoturbulenceinrapidlyrotatingcore-collapsesupernovae,Halevi_Moesta__2018__mnras__r-Processnucleosynthesisfromthree-dimensionaljet-drivencore-collapsesupernovaewithmagneticmisalignments,Moesta_et_al__2018__apj__r-processNucleosynthesisfromThree-dimensionalMagnetorotationalCore-collapseSupernovae,Bugli__2020__MonthlyNoticesoftheRoyalAstronomicalSociety__TheImpactofNonDipolarMagneticFieldsinCoreCollapseSupernovae,Kuroda_et_al__2020__apj__MagnetorotationalExplosionofaMassiveStarSupportedbyNeutrinoHeatinginGeneralRelativisticThreeDimensionalSimulations,Aloy__2021__MonthlyNoticesoftheRoyalAstronomicalSociety__MagnetorotationalCoreCollapseofPossibleGRBProgenitorsII.FormationofProtomagnetarsandCollapsars,Grimmett__2021__MonthlyNoticesoftheRoyalAstronomicalSociety__TheChemicalSignatureofJetDrivenHypernovae,Obergaulinger__2021__mnras__MagnetorotationalCoreCollapseofPossibleGRBProgenitorsIII.ThreeDimensionalModels,Bugli__2021__mnras__ThreeDimensionalCoreCollapseSupernovaewithComplexMagneticStructuresI.ExplosionDynamics,Reichert__2021__MonthlyNoticesoftheRoyalAstronomicalSociety__NucleosynthesisinMagnetoRotationalSupernovae,Varma__2021__MonthlyNoticesoftheRoyalAstronomicalSociety__AComparisonof2DMagnetohydrodynamicSupernovaSimulationswiththeCOCONUTFMTandAENUSALCARCodes,Obergaulinger__2022__MonthlyNoticesoftheRoyalAstronomicalSociety__MagnetorotationalCoreCollapseofPossibleGammaRayBurstProgenitorsIV.aWiderRangeofProgenitors}.
This set of simulations is not homogeneous in terms of the input
physics and numerical methods (see Fig.~\ref{fig:overview_studies} for an overview of the methods used in several studies) and shows a clear tendency toward more
realistic modeling during the past decade.  The detailed
nucleosynthesis has been computed only for a limited
subset of them.  Even in cases where no nucleosynthesis calculations are available, the (M)HD simulation results
allow for an assessment of the thermodynamic properties of the ejecta
that determine which nuclear reactions can take place. 

The structure of the core sets the conditions for the possible
formation and propagation of jets.  While an accretion disk is unlikely
to form during this phase, the rotation may be fast enough to induce a
moderate anisotropy of the PNS and the neutrino emission.  In the
absence of magnetic fields, a delayed neutrino-driven explosion may
eject gas in an anisotropic, albeit not collimated, geometry.  Jets
can form if, additionally, the PNS possesses a magnetic
field with a strong large-scale, e.g., dipole, component.  While the
details may vary, the evolution tends to proceed similarly to the
model of magnetic towers.  

Magnetorotational jets may initiate a prompt explosion whose kinetic energies reach the values typical for HNe.  For weaker fields and slower rotation, the standard neutrino-heating mechanism may launch an explosion, after which energetically subdominant jets may be launched into the hot bubble left behind by the expanding SN shock wave.

Since these jets are launched near the deleptonized surface layers of
the PNS, matter at their base is neutron-rich.  If it is accelerated
by the magnetic field very efficiently to high velocities, the gas
will cross the region in which it is exposed to the intense neutrino
radiation emanating from the PNS sufficiently rapidly to avoid an
increase of $Y_e$ to its equilibrium value close to $Y_e \simeq 0.5$ (for typical neutrino properties found in (M)HD simulations).  As a
consequence, parts of the jet, in particular near its head, can
maintain $Y_e \sim 0.2$, which makes them promising sites of the
r-process.  Nucleosynthesis calculations like the ones by
\cite{Nishimura2006,Tominaga__2009__apj__Aspherical_Properties_of_Hydrodynamics_and_Nucleosynthesis_in_Jet-Induced_Supernovae,Winteler_et_al__2012__apjl__MagnetorotationallyDrivenSupernovaeastheOriginofEarlyGalaxyr-processElements,Nishimura_et_al__2015__apj__Ther-processNucleosynthesisintheVariousJet-likeExplosionsofMagnetorotationalCore-collapseSupernovae,Nishimura_et_al__2017__apjl__TheIntermediateRProcessinCoreCollapseSupernovaeDrivenbytheMagnetoRotationalInstability,Halevi_Moesta__2018__mnras__r-Processnucleosynthesisfromthree-dimensionaljet-drivencore-collapsesupernovaewithmagneticmisalignments,Moesta_et_al__2018__apj__r-processNucleosynthesisfromThree-dimensionalMagnetorotationalCore-collapseSupernovae,Reichert__2021__MonthlyNoticesoftheRoyalAstronomicalSociety__NucleosynthesisinMagnetoRotationalSupernovae}
support this proposition, demonstrating at least weak r-process patterns reaching the second r-process peak, in some cases even a full r-process up to the actinides.  
Fig.~\ref{fig:Moesta18Abund} shows results for the dynamical evolution (top panels) and the nucleosynthesis (bottom panels) of cores with very strong pre-collapse field strength \citep{Moesta_et_al__2018__apj__r-processNucleosynthesisfromThree-dimensionalMagnetorotationalCore-collapseSupernovae}.  All explosions develop jets, but non-axisymmtric instabilities can lead to large perturbations and convert the jet into a wider polar outflow (right column).  The nucleosynthesis pattern depends on assumptions for the neutrino luminosities the  tracer particles are exposed to.  The r-process is most successful, reaching the 3rd peak, for the combination of the strongest fields and thus fastest, most collimated jets with the lowest neutrino luminosities.

\begin{figure}
  \centering
  \includegraphics[width=\linewidth]{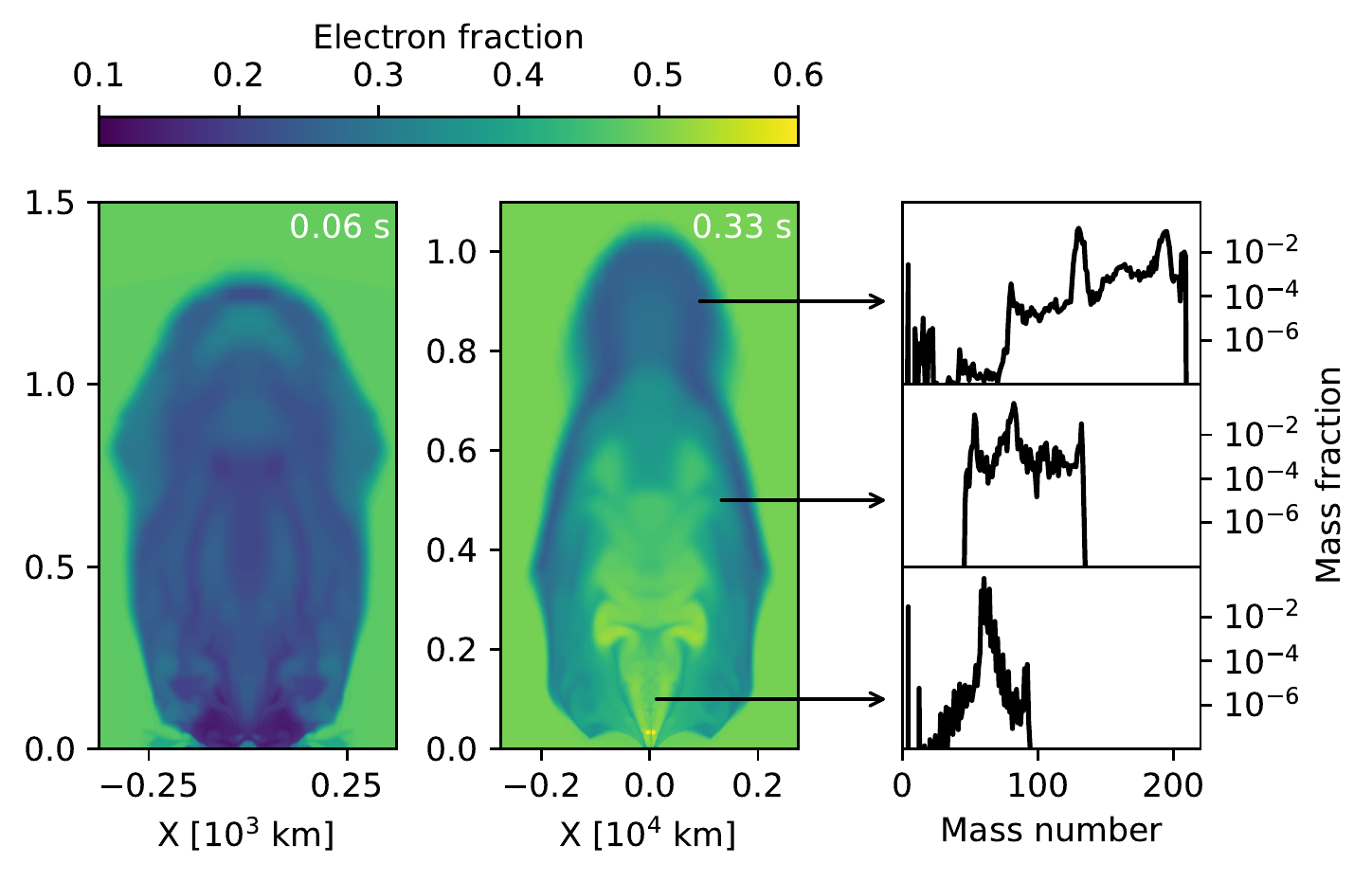}
  \caption{Electron fraction for a strongly magnetized MR-SNe model (35OC-Rs of  \citealt{Obergaulinger__2021__mnras__MagnetorotationalCoreCollapseofPossibleGRBProgenitorsIII.ThreeDimensionalModels,Aloy__2021__MonthlyNoticesoftheRoyalAstronomicalSociety__MagnetorotationalCoreCollapseofPossibleGRBProgenitorsII.FormationofProtomagnetarsandCollapsars,Reichert__2021__MonthlyNoticesoftheRoyalAstronomicalSociety__NucleosynthesisinMagnetoRotationalSupernovae}) shortly after core bounce ($t_{pb}=0.06\,\mathrm{s}$, left panel) and for a later time ($t_{pb}=0.33\,\mathrm{s}$, middle panel). Examples of synthesized nuclei in different parts of the jet are shown in the right panels. The panels show mass number versus mass fractions.
  }
  \label{Fig:jet_nucleosynthesis}
\end{figure}

Series of models using the same input physics and progenitor, but varying the initial magnetic field \citep[e.g.,][]{Reichert__2021__MonthlyNoticesoftheRoyalAstronomicalSociety__NucleosynthesisinMagnetoRotationalSupernovae} suggest a continuum of nucleosynthesis results from weakly magnetized cores producing only few elements beyond the Fe group to a full r-process pattern for the strongest magnetic fields. As the nucleosynthetic yields depend on an interplay of magnetic and neutrino pressure, similar results are obtained when varying the neutrino flux and keeping the magnetic field strengths constant (e.g., \citealt{Nishimura_et_al__2017__apjl__TheIntermediateRProcessinCoreCollapseSupernovaeDrivenbytheMagnetoRotationalInstability}).
Even within the same jet, these dependencies lead to a highly non-uniform composition.  The electron fraction of the ejecta can vary strongly with time.  In the case of a prompt jet shown in Fig.~\ref{Fig:jet_nucleosynthesis}, neutron-rich matter is ejected during the first few ms, whereas later ejecta are more proton-rich.  Thus, the front regions of the jet have lower $Y_e \gtrsim 0.2$, which allows for the synthesis of nuclei up to $A \approx 200$.  In the middle and rear sections of the jet, the nucleosynthesis pattern is more restricted with a cut-off at lower mass numbers.

These results should be taken with a grain of salt, though. A few caveats apply:
\begin{itemize}
\item The highest mass numbers are reached by those numerical models
  that employ the most significant simplifications in the neutrino
  physics.  Simulations with state-of-the-art two-moment neutrino
  transport tend to show ejecta with higher (i.e., less neutron-rich) than ones with more
  approximate methods such as leakage schemes. Some studies partly account for this well known effect by artificially varying the neutrino luminosities or the electron fraction \citep[e.g.,][]{Winteler_et_al__2012__apjl__MagnetorotationallyDrivenSupernovaeastheOriginofEarlyGalaxyr-processElements, Nishimura_et_al__2017__apjl__TheIntermediateRProcessinCoreCollapseSupernovaeDrivenbytheMagnetoRotationalInstability,Moesta_et_al__2018__apj__r-processNucleosynthesisfromThree-dimensionalMagnetorotationalCore-collapseSupernovae}. However, this prevents a consistent view on the synthesis of heavy r-process elements.
\item Near the jet base, the helical field lines along which the gas
  moves can have a low pitch. Consequently, the propagation along the
  jet axis can be slow compared to the rotation around it.  For such a
  geometry, three-dimensional models
  \citep{Obergaulinger__2021__mnras__MagnetorotationalCoreCollapseofPossibleGRBProgenitorsIII.ThreeDimensionalModels}
  show that before being ejected in the jet, fluid elements may spend
  more time in the vicinity of the PNS than is needed by neutrino
  reactions to increase $Y_e$ and thus reduce the prospects for the r-process.
\item Another effect that depends on the geometry and strength of the
  magnetic field is the possible development of non-axisymmetric
  instabilities in the jet after its formation.
  \cite{Mosta_et_al__2014__apjl__MagnetorotationalCore-collapseSupernovaeinThreeDimensions}
  found that kink modes can destabilize jets to the point of
  converting them from a collimated geometry to wide outflows with
  slower outward propagation and quite different nucleosynthesis
  conditions.  Not all similar studies have found a similarly
  pronounced effect on the jet
  \citep{Kuroda_et_al__2020__apj__MagnetorotationalExplosionofaMassiveStarSupportedbyNeutrinoHeatinginGeneralRelativisticThreeDimensionalSimulations,Obergaulinger__2021__mnras__MagnetorotationalCoreCollapseofPossibleGRBProgenitorsIII.ThreeDimensionalModels}
  and the physical and numerical conditions for the instability remain
  an open issue.
\item Very strong fields with a large-scale geometry are required for
  generating very energetic jets.  Unless a suitable magnetic field is
  present already before collapse, it has to develop after bounce by,
  e.g., a dynamo based on the MRI or convection. This process may
  under favorable conditions be very rapid
  \citep{Moesta_et_al__2015__nat__Alarge-scaledynamoandmagnetoturbulenceinrapidlyrotatingcore-collapsesupernovae},
  but usually can take many rotational or dynamical timescales
  \citep{Raynaud__2020__ScienceAdvances__MagnetarFormationthroughaConvectiveDynamoinProtoneutronStars,ReboulSalze__2021__aap__AGlobalModeloftheMagnetorotationalInstabilityinProtoneutronStars,Raynaud__2022__MonthlyNoticesoftheRoyalAstronomicalSociety__GravitationalWaveSignatureofProtoNeutronStarConvectionI.MHDNumericalSimulations,White__2022__TheAstrophysicalJournal__OntheOriginofPulsarandMagnetarMagneticFields},
  if it does not fail to reach sufficiently strong magnetic energy
  altogether.  Hence, the parameter space for jets that can lead to
  the r-process may be a narrow one.
\item A non dipolar magnetic field or a misalignment with respect to the rotational axis will lead also to a slower or an off-axis development of the jet \citep[e.g.,][]{Halevi_Moesta__2018__mnras__r-Processnucleosynthesisfromthree-dimensionaljet-drivencore-collapsesupernovaewithmagneticmisalignments,Bugli__2020__MonthlyNoticesoftheRoyalAstronomicalSociety__TheImpactofNonDipolarMagneticFieldsinCoreCollapseSupernovae,Bugli__2021__mnras__ThreeDimensionalCoreCollapseSupernovaewithComplexMagneticStructuresI.ExplosionDynamics}. \citet{Halevi_Moesta__2018__mnras__r-Processnucleosynthesisfromthree-dimensionaljet-drivencore-collapsesupernovaewithmagneticmisalignments} showed that this slower jet expansion causes less neutron-rich conditions and therefore a reduction in ejected heavy elements.
\end{itemize}

\begin{figure}
 \includegraphics[width=\columnwidth]{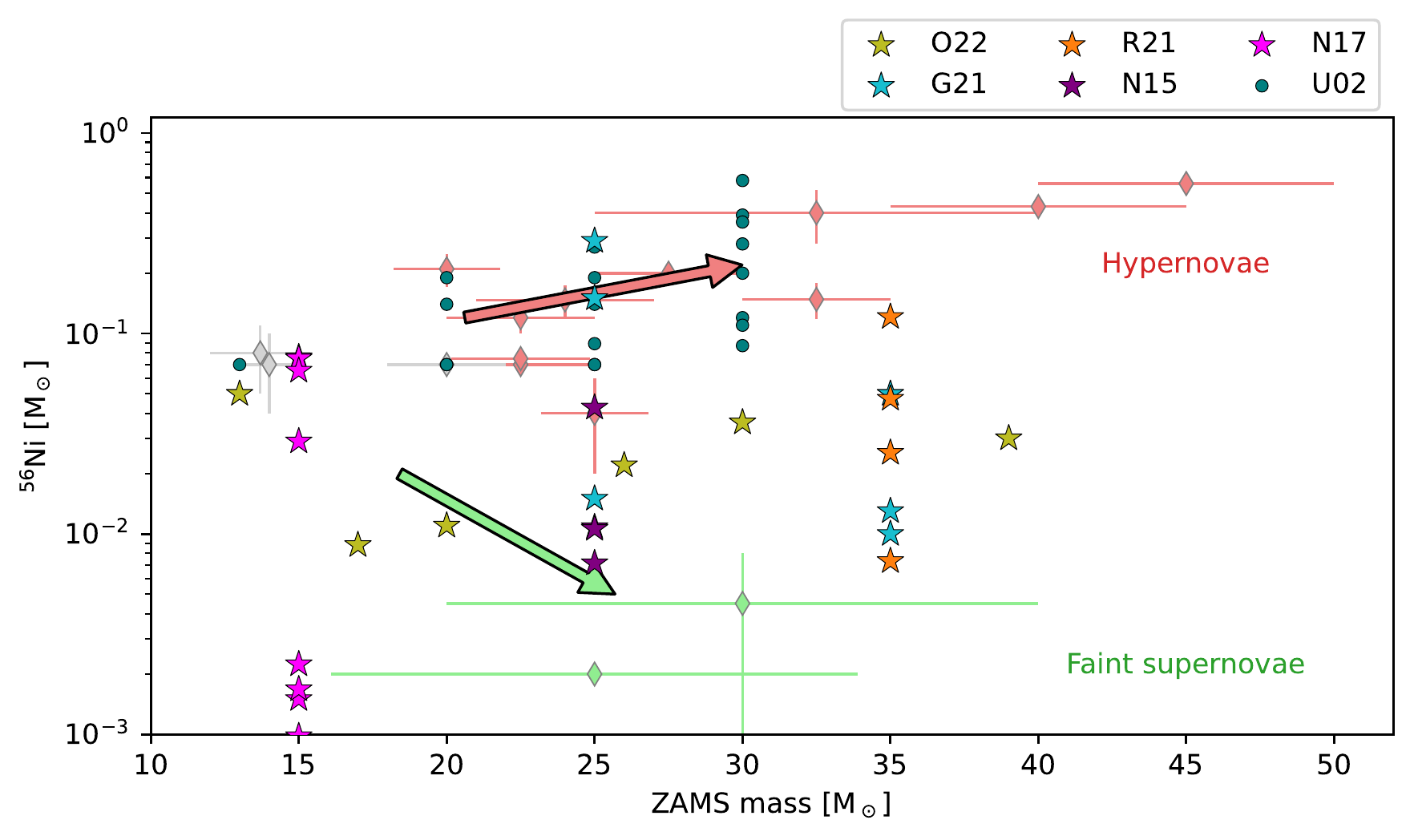}\\
 \caption{Observed Nickel masses versus ZAMS masses (diamonds). Red colors indicate hypernovae with kinetic energies $E_\mathrm{kin}\ge 10^{52}\,\mathrm{erg}$, green colors indicate faint supernovae \citep[taken from][]{Bufano2012,Nomoto_et_al__2013__araa__NucleosynthesisinStarsandtheChemicalEnrichmentofGalaxies,DElia2015}. Colored stars show theoretical predictions from 2D simulations of \citet[][ yellow]{Obergaulinger__2022__MonthlyNoticesoftheRoyalAstronomicalSociety__MagnetorotationalCoreCollapseofPossibleGammaRayBurstProgenitorsIV.aWiderRangeofProgenitors}, \citet[][cyan]{Grimmett__2021__MonthlyNoticesoftheRoyalAstronomicalSociety__TheChemicalSignatureofJetDrivenHypernovae}, \citet[][orange]{Reichert__2021__MonthlyNoticesoftheRoyalAstronomicalSociety__NucleosynthesisinMagnetoRotationalSupernovae}, \citet[][purple]{Nishimura_et_al__2015__apj__Ther-processNucleosynthesisintheVariousJet-likeExplosionsofMagnetorotationalCore-collapseSupernovae}, and \citet[][magenta]{Nishimura_et_al__2017__apjl__TheIntermediateRProcessinCoreCollapseSupernovaeDrivenbytheMagnetoRotationalInstability}. For comparison, teal circles show predictions from the mixing-fallback model of \citet{Umeda2002}.  }
 \label{fig:nickel_prod}
\end{figure}

The high explosion energies found for strongly magnetized cores can
reach the range of HNe.  Observationally, this class of explosions is
connected to the production of a large mass of $\AZNucleus{56}{}{Ni}$
of at least $0.1 \, \msol$ \citep{Nomoto__2006__NuclearPhysicsA__NucleosynthesisYieldsofCoreCollapseSupernovaeandHypernovaeandGalacticChemicalEvolution,Nomoto_et_al__2013__araa__NucleosynthesisinStarsandtheChemicalEnrichmentofGalaxies}.
A viable explanation for HNe should be consistent with both findings.
In regular CCSNe, the amount of ejecta fulfilling the requirements for
the synthesis of $\AZNucleus{56}{}{Ni}$, i.e., high temperatures of $T
\ge 4 \, \mathrm{GK}$ in symmetric or proton-rich conditions \citep[e.g.,][]{Magkotsios2011}, tends to
increase with the explosion energy as a more energetic shock wave can
maintain high temperatures for longer.  It is, however, difficult for
models of regular, neutrino-driven CCSNe to reproduce HN-like
energies.  Magnetorotational models, on the other hand, quite commonly
reach them, which prompts the question of whether they can produce
enough $^{56}$Ni.
\cite{Suwa__2015__MonthlyNoticesoftheRoyalAstronomicalSociety__HowMuchCan56NiBeSynthesizedbytheMagnetarModelforLongGammaRayBurstsandHypernovae}
explored the available parameter space, finding HN-like $^{56}$Ni masses for
cores with the fastest rotating and strongest magnetized PNSs.
Simulations with simplified neutrino transport
\cite{Winteler_et_al__2012__apjl__MagnetorotationallyDrivenSupernovaeastheOriginofEarlyGalaxyr-processElements,Mosta_et_al__2014__apjl__MagnetorotationalCore-collapseSupernovaeinThreeDimensions,Nishimura_et_al__2015__apj__Ther-processNucleosynthesisintheVariousJet-likeExplosionsofMagnetorotationalCore-collapseSupernovae,Nishimura_et_al__2017__apjl__TheIntermediateRProcessinCoreCollapseSupernovaeDrivenbytheMagnetoRotationalInstability,Halevi_Moesta__2018__mnras__r-Processnucleosynthesisfromthree-dimensionaljet-drivencore-collapsesupernovaewithmagneticmisalignments,Moesta_et_al__2018__apj__r-processNucleosynthesisfromThree-dimensionalMagnetorotationalCore-collapseSupernovae} 
generate jets that might be too neutron-rich, while simulations with
M1 neutrino transport show jets with higher $Y_e$
\citep{Obergaulinger_Aloy__2020__mnras__MagnetorotationalCoreCollapseofPossibleGRBProgenitorsIExplosionMechanisms,Obergaulinger__2021__mnras__MagnetorotationalCoreCollapseofPossibleGRBProgenitorsIII.ThreeDimensionalModels},
though even in such cases the total $^{56}$Ni masses fall shy of the values
for HNe
\citep{Reichert__2021__MonthlyNoticesoftheRoyalAstronomicalSociety__NucleosynthesisinMagnetoRotationalSupernovae,Obergaulinger__2022__MonthlyNoticesoftheRoyalAstronomicalSociety__MagnetorotationalCoreCollapseofPossibleGammaRayBurstProgenitorsIV.aWiderRangeofProgenitors}.
\cite{Grimmett__2021__MonthlyNoticesoftheRoyalAstronomicalSociety__TheChemicalSignatureofJetDrivenHypernovae}
explored the production of $^{56}$Ni in a set of jet-driven models with
parameterized explosion energy and electron fraction. They found that,
taking together the production in the jet and in the more spherical
off-axis ejecta, $^{56}$Ni masses of up to $0.45 \, \msol$ can be reached for
very energetic explosions.  Such values would be consistent with HNe
of type Ic-BL.  When interpreting results from other models, it should
be taken into account that the formation of $^{56}$Ni can continue for a
longer time than what can be currently covered by simulations, in
particular if expensive methods for the neutrino transport are used.
Fig.~\ref{fig:nickel_prod} summarises the $^{56}$Ni production found in several numerical studies and compares the results to observed events from faint CCSNe to HNe. Between the HNe and faint SNe branch, there is a continuous spectra of observations, not shown in the figure \citep[see, e.g.,][]{Hamuy2003}. While no final answer can be given, numerical models seem able to reproduce at least partially the observed range.

The aforementioned limitation of the achievable explosion energy in
regular CCSNe prompted interest in an alternative explosion scenario
in which jets are predominant
\citep{Soker_et_al__2013__AstronomischeNachrichten__Thejetfeedbackmechanism(JFM):Fromsupernovaetoclustersofgalaxies,Papish__2014__MonthlyNoticesoftheRoyalAstronomicalSociety__ExplodingCoreCollapseSupernovaebyJetsDrivenFeedbackMechanism}.
Three-dimensional simulations of CCSNe show that turbulent flows, in
particular spiral modes of the SASI, in the gas accreted onto the
center can lead to the accumulation of angular momentum of
stochastically varying magnitude and direction around the PNS,
potentially up to the point of the formation of an accretion disk.
From this configuration, jets may be launched along the rotational
axis.  These so-called jittering jets should be highly unsteady, with
their mass and energy fluxes as well as propagation directions
changing according to the random conditions at the center.  Each of
the jets would only be weak compared to the ones described previously
and, thus, unable to initiate a jet-like explosion.  Instead, they are
stopped behind the stalled CCSN shock wave, where their energy is
dissipated.  This sequence of events constitutes an additional way of
transmitting energy from the center to the shock ways, working in
parallel to neutrino heating.
\cite{Soker_et_al__2013__AstronomischeNachrichten__Thejetfeedbackmechanism(JFM):Fromsupernovaetoclustersofgalaxies} and \cite{Papish__2014__MonthlyNoticesoftheRoyalAstronomicalSociety__ExplodingCoreCollapseSupernovaebyJetsDrivenFeedbackMechanism}
posit that the feedback between accretion and jet emission should be
active under quite general conditions and finally trigger shock
revival, at which point the decrease of the accretion rate would shut
down the jet formation.
\cite{Papish__2015__MonthlyNoticesoftheRoyalAstronomicalSociety__ACallforaParadigmShiftfromNeutrinoDriventoJetDrivenCoreCollapseSupernovaMechanisms} and \cite{Soker__2017__TheAstrophysicalJournal__MagnetarPoweredSuperluminousSupernovaeMustFirstBeExplodedbyJets}
argue that it can lead to very high explosion energies.  While several
variations of this scenario have been proposed, a confirmation in
self-consistent numerical models and an exploration of the
nucleosynthesis in this kind of explosion are still pending.

\subsection{Long GRBs}
\label{sSek:GRBs}

The jets of long GRBs are characterized by very high Lorentz factors
and small opening angles.  Strictly speaking, these properties have been inferred for the phase at which the $\gamma$/X-ray radiation is emitted, which is after the breakout into the circumstellar medium.  How they are related to the conditions in the central engine where the jets originate
and where most of the nucleosynthesis occurs is a matter of ongoing
research.  If the engine is a PM, the basic processes are similar to
those in strongly magnetized and rapidly rotating CCSNe, though
operating at a more extreme level and, possibly, at later times after
a CCSN has set in.  For the BH in a collapsar, on the other hand, the
presence of a BH-accretion disk system allows for different
mechanisms.  

Numerical studies exploring the dynamics and nucleosynthesis of GRBs
of both classes tend to differ from those targeting CCSNe in various
aspects.  To achieve the required longer simulation times of the order
of several seconds, compromises are often made in the treatment of
neutrinos.  For the same reason, but in the case of a BH as central
object also out of necessity, many models excise the innermost, at
least, tens of km and replace them with an external gravitational
potential or metric and an inner boundary condition \citep[][]{MacFadyen_Woosley__1999__ApJ__Collapsar}.  This method
leaves the choice of the initial and boundary conditions open.  The
former are commonly adapted from the profiles of CCSN progenitors,
often without accounting for the changes due to the collapse and
immediate post-bounce evolution.  Many studies consider the evolution
of jets injected via a nozzle with a given opening angle around the
rotational axis at which a flow with prescribed density, composition,
entropy, Lorentz factor, and magnetization is injected.  In other
cases, the jet acceleration can be followed explicitly as the MHD
stresses or neutrino heating are included.

\begin{figure}
    \centering
    \includegraphics[width=0.95\columnwidth]{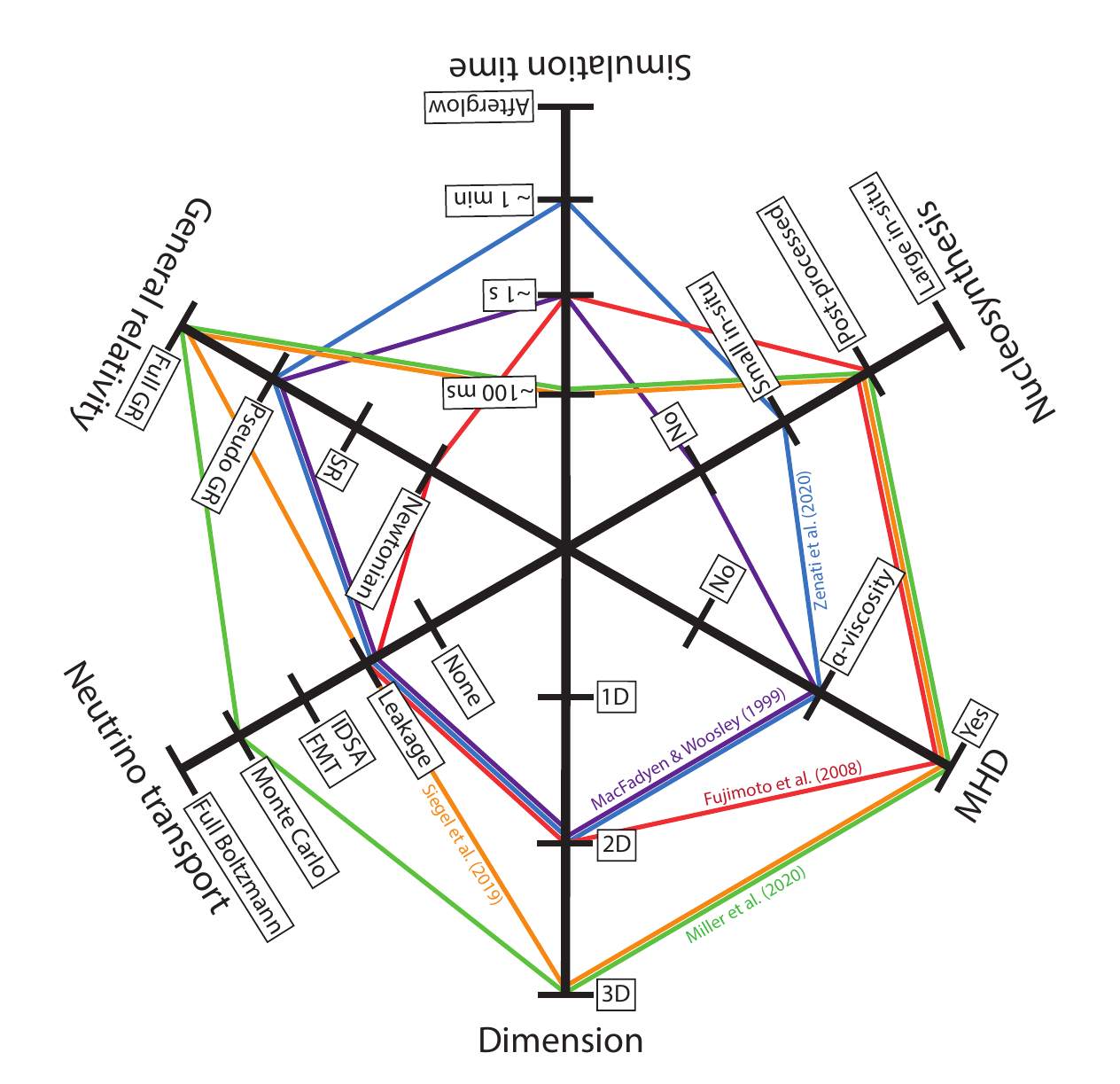}
    \caption{Overview of chosen physical inputs within collapsar simulations. Going inwards on the axis indicates a more approximate approach. Shown are the works of \citet[][green]{Miller__2020__TheAstrophysicalJournal__FullTransportGeneralRelativisticRadiationMagnetohydrodynamicsforNucleosynthesisinCollapsars}, \citet[][orange]{Siegel__2019__Nature__CollapsarsAsaMajorSourceofRProcessElements}, \citet[][red]{Fujimoto__2008__TheAstrophysicalJournal__NucleosynthesisinMagneticallyDrivenJetsfromCollapsars}, \citet[][blue]{Zenati__2020__MonthlyNoticesoftheRoyalAstronomicalSociety__NuclearBurninginCollapsarAccretionDiscs}, and \citet[][purple]{MacFadyen_Woosley__1999__ApJ__Collapsar}.}
    \label{fig:collapsar_literature}
\end{figure}

While neutrinos may contribute to a minor degree to accelerating the
outflows, the main energy source of a PM-driven jet is the rotational
energy extracted from the (millisecond) PNS by its extremely strong
magnetic fields.  Studies of PMs \citep[e.g.,][]{Metzger2018a}
commonly include the spin-down of the PM by a magnetic field, usually
assumed to be of dipole geometry, potentially modified by the presence
of fall-back accretion, and a wind driven by the radiation pressure of
the neutrinos emanating from the PM
\citep[][]{Qian__1996__TheAstrophysicalJournal__NucleosynthesisinNeutrinoDrivenWinds.I.thePhysicalConditions}.
For these assumptions to apply, the PM should be surrounded by a
low-density medium, which can be achieved by a successful
(magnetorotational) CCSN having cleared out the core before the PM
activity sets in.  In this case, the outflows generated by the PM
would be faster than the CCSN shock or the magnetorotationally driven
CCSN shock and overtake them during the propagation to the
surface or after breakout from the surface.  Calculations of the
nucleosynthesis of these events should take into account the
contributions of the jets and the more spherical winds.

As in the case of regular and magnetorotational CCSNe, for PM-driven
GRBs a particular focus lies on the production of
$\AZNucleus{56}{}{Ni}$ and of heavy (r-process) nuclei.  The
simplified models of
\cite{Metzger_et_al__2007__apj__Proto-NeutronStarWindswithMagneticFieldsandRotation}
showed that, in contrast to the case of regular PNS winds, the
required conditions for the latter, fast ejection and high entropies
and/or low electron fraction, can be met by the wind of a strongly
magnetized, rapidly rotating PM.
\cite{Vlasov_et_al__2014__mnras__Neutrino-heatedwindsfromrotatingprotomagnetars}
and the follow-up study of
\cite{Vlasov__2017__MonthlyNoticesoftheRoyalAstronomicalSociety__NeutrinoHeatedWindsfromMillisecondProtomagnetarsAsSourcesoftheWeakRProcess}
built upon this result, analyzing steady-state models of winds driven
by neutrino emission and shaped by the strong dipolar magnetic field
of the PM.  The calculations follow the matter ejected from the polar
region of the PM along open field lines and its evolution under the
influence of neutrino reactions.  The parameters of the model include
the mass, rotation, magnetization, and neutrino luminosities of the PM.  For
rotation periods around a few ms, the study estimates an amount of up
to $\sim 10^{-4} \, \msol$ of r-process ejecta per event, which would
make PM-driven GRBs very relevant sources of the total galactic mass
of r-process elements up to $A \sim 90 - 120$.  The heaviest elements
are usually produced off-axis at the outer edge of the cone of open
field lines.  The final opening angle of this component of the outflow
will depend on the effectiveness of jet collimation in the actual
post-collapse stellar (density) profiles.  The dependence of the final
yields on the PM parameters allows to account for the diversity of
observed events.  The tendency for faster rotation reaching higher
atomic masses, at least if the ejecta are moderately neutron-rich,
was confirmed by
\cite{Ekanger__2022__MonthlyNoticesoftheRoyalAstronomicalSociety__SystematicExplorationofHeavyElementNucleosynthesisinProtomagnetarOutflows}.  

\cite{Thompson__2018__MonthlyNoticesoftheRoyalAstronomicalSociety__HighEntropyEjectionsfromMagnetizedProtoNeutronStarWindsImplicationsforHeavyElementNucleosynthesis}
performed axisymmetric MHD simulations with approximate neutrino
heating of the winds of strongly magnetized, yet non-rotating PNSs.
They pointed out the importance of the variability of the flow, which
could not be captured by steady-state models.  At low latitudes, fluid
elements can be trapped in regions of closed fields lines, from which
plasmoids can form and erupt.  After this process, the closed
magnetosphere reforms due to magnetic reconnection.  as long as the
gas is trapped, neutrino heating increases its entropy to very high
values.  In combination with the fast expansion in the explosive
plasmoid ejection events, the conditions for the r-process may be met.
While each such event only ejects little mass, the mechanism can
repeat.  A total r-process mass of $\sim 10^{-5} \, \msol$ may be
obtained.  This number might be modified for rapidly rotating PNSs,
which would help to alleviate possible conflicts with the observed
birth rates of magnetars.  If the matter is proton-rich, the $\nu$p
process might work instead.

Many numerical models of collapsars follow the technique of the
pioneering study of \cite{MacFadyen_Woosley__1999__ApJ__Collapsar} and
evolve a BH, represented by an inner boundary
condition, and an accretion disk at the center of a massive star with
relativistic MHD and approximate neutrino physics (see Fig.~\ref{fig:collapsar_literature} for an overview of numerical methods).  Basic elements of
the dynamics have been established.  For sufficiently high specific
angular momentum, the matter falling toward the newly formed BH
accumulates in a disk or torus in the equatorial region.  Accretion
onto the BH is regulated by the transport of angular momentum in the
disk.  While this process is most likely ultimately due to turbulence
driven by the MRI, it can be described by an effective viscosity.  It
is accompanied by heating of the gas, which may cause an outflow from
the disk.  The viscous outflow can carry an energy on the level of a
HN.  Additionally, a GRB can be generated by a jet launched due to
neutrino heating in an evacuated funnel along the rotational axis or
the magnetic fields of the BH-disk system.
Within this class of models, \cite{Tominaga__2007__TheAstrophysicalJournal__TheConnectionbetweenGammaRayBurstsandExtremelyMetalPoorStarsBlackHoleFormingSupernovaewithRelativisticJets} found that explosions with energetic jets are compatible with GRBs, the observational signature of HNe, and yields of heavy elements like the ones observed in extremely metal-poor stars.
\cite{Fujimoto__2008__TheAstrophysicalJournal__NucleosynthesisinMagneticallyDrivenJetsfromCollapsars}
showed the formation of a solar r-process pattern for stars with fast,
though not extreme, rotation and very strong magnetic fields as well
as the production of large quantities of $^{56}$Ni.  The later models of
\cite{Ono__2009__ProgressofTheoreticalPhysics__ExplosiveNucleosynthesisinMagnetohydrodynamicalJetsfromCollapsars,Ono__2012__ProgressofTheoreticalPhysics__ExplosiveNucleosynthesisinMagnetohydrodynamicalJetsfromCollapsars.IIHeavyElementNucleosynthesisofSPRProcesses,Nakamura__2015__aap__RProcessNucleosynthesisintheMHDneutrinoHeatedCollapsarJet}
confirm the basic picture and also shed light on contributions by
other processes.

Parts of the fluid elements ejected via the wind or the jet originate
from regions where the densities and temperatures are sufficient for
NSE.  As
\cite{Surman__2004__TheAstrophysicalJournal__NeutrinosandNucleosynthesisinGammaRayBurstAccretionDisks,McLaughlin__2005__NuclearPhysicsA__ProspectsforObtaininganRProcessfromGammaRayBurstDiskWinds,Surman__2006__TheAstrophysicalJournal__NucleosynthesisintheOutflowfromGammaRayBurstAccretionDisks}
demonstrated, whether or not $\nu_e$ or $\bar{\nu}_e$ are trapped sets
the electron fraction of the accretion disks and thus the possible
nucleosynthesis products in outflows emanating from them.  At the
highest accretion rates, disks are optically thick to both species of
neutrinos.  The result is neutron-rich matter, thus allowing for the
r-process.  The lowest mass accretion rates and, thus, optically thin
conditions restrict the nucleosynthesis to Fe group and lighter elements.

\begin{figure}
    \centering
    \includegraphics[width=1\columnwidth]{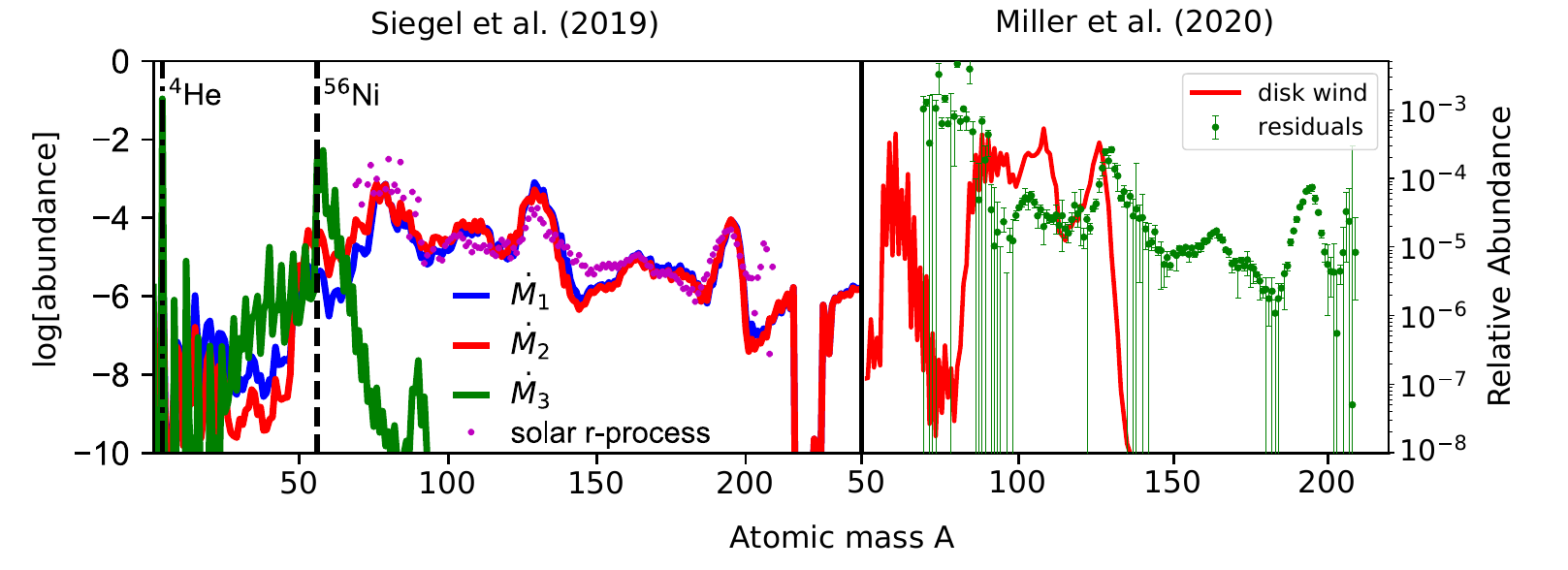}
    \caption{%
        Final abundances in collapsar simulations of     \cite{Siegel__2019__Nature__CollapsarsAsaMajorSourceofRProcessElements} (left) and     \cite{Miller__2020__TheAstrophysicalJournal__FullTransportGeneralRelativisticRadiationMagnetohydrodynamicsforNucleosynthesisinCollapsars} (right).
        In the left panel, line colors denote models with different assumptions for the mass accretion rate, $\dot{M}$, onto the central BH ($\dot{M}_1 > \dot{M}_2 > \dot{M}_3$).  Symbols show the solar r-process pattern.
        In the right panel, the red line represents the nucleosynthesis in the disk winds and the green symbols are the solar r-process residuals. Figure adapted with permission of the authors and the publishers from \cite{Siegel__2019__Nature__CollapsarsAsaMajorSourceofRProcessElements} (left, \copyright~ ) Springer Nature) and \cite{Miller__2020__TheAstrophysicalJournal__FullTransportGeneralRelativisticRadiationMagnetohydrodynamicsforNucleosynthesisinCollapsars} (right, \copyright~ IOP Publishing).
    }
    \label{fig:CollapsarAbund}
\end{figure}

Based on three-dimensional GRMHD simulations set up as representative
models for BH-disk systems at various stages of the evolution of a
collapsar,
\cite{Siegel__2019__Nature__CollapsarsAsaMajorSourceofRProcessElements}
argue for collapsars as a main site of the r-process.  When the
accretion rate is very high, the disk forms an optically thick
neutrino-dominated accretion flow or NDAF, in which electrons are
degenerate and neutrinos are in local thermodynamic equilibrium with
matter.  Consequently, neutrino cooling is fast and dominates the
dynamics.  Furthermore, the frequent emission and absorption of
neutrinos drives the electron fraction to an equilibrium value that
can be as low as $Y_e$ at the surface of a PNS \citep[][discussed the
same effects, though with a focus on BH-disk systems in binary
mergers]{Just__2022__MonthlyNoticesoftheRoyalAstronomicalSociety__NeutrinoAbsorptionandOtherPhysicsDependenciesinNeutrinoCooledBlackHoleAccretionDiscs}.
As long as these conditions hold, gas ejected from the disk can
therefore be very neutron-rich, making such outflows possible
candidates for the r-process up to the 3rd peak.  The absence of a
central neutrino source like a PNS/PM reduces the possibility of
neutrino interactions raising $Y_e$ during the propagation of the
wind.  At lower accretion rates, the disk becomes less dense and
ultimately transparent to neutrinos.  From this point onward,
neutrinos and matter decouple and the disk matter is less neutron
rich.  With the dynamics dominated by MHD or viscous stresses,
outflows keep being emitted, but their composition will favour the
production of Fe group and, for the lowest accretion rate, light
elements.  Synthesizing high masses of r-process material, collapsars
should play an important, possibly dominant, role in the production of
the heaviest elements in the Galaxy \cite[see
also][]{Siegel__2019__EuropeanPhysicalJournalA__GW170817theFirstObservedNeutronStarMergerandItsKilonovaImplicationsfortheAstrophysicalSiteoftheRProcess}.
Working in a similar framework as
\cite{Siegel__2019__Nature__CollapsarsAsaMajorSourceofRProcessElements},
but employing a more sophisticated neutrino transport method,
\cite{Miller__2020__TheAstrophysicalJournal__FullTransportGeneralRelativisticRadiationMagnetohydrodynamicsforNucleosynthesisinCollapsars}
found somewhat less neutron-rich conditions and, thus, ejecta that
only reach the 2nd r-process peak (Fig.~\ref{fig:CollapsarAbund} shows a comparison of the results of these two studies). 

\cite{Zenati__2020__MonthlyNoticesoftheRoyalAstronomicalSociety__NuclearBurninginCollapsarAccretionDiscs}
considered nuclear burning in collapsar disks.  If disks form with
high masses and densities, a detonation is possible, leading to the
release of up to $\sim 5 \times 10^{50} \, \erg$ and $\sim 0.1 \,
\msol$ of $^{56}$Ni.  While these values are not enough to account for the
HNe associated with long GRBs, they could provide an additional
component.

As for CCSNe with jets, further improvements of the numerical methods
are desirable.  Combining accurate neutrino physics and
high-resolution MHD simulations in a three-dimensional geometry in
simulations covering evolution times of many seconds is a major
challenge.  With current numerical codes and computational resources,
the choice is between including all relevant physical effects and
running for long times.  While the drawbacks of choosing the latter are
obvious, a main limitation of the former option is that it requires
relying on simplified, possibly artificial initial conditions for
simulations that are set in relatively late phases of the global evolution.

\section{Observational constraints}

Whether or not jet-SNe are able to provide the necessary conditions to host an r-process is still under debate. The previous sections only tackled this question from a modelling perspective, however, some hints can also be obtained by looking at observational constraints. 

When a star forms it is made up from matter of its surrounding dust cloud. As a consequence, it locks a snapshot of the current chemical composition in its atmosphere. Very old stars are therefore witnesses of an ancient composition and the abundances can be determined by observing their spectra. A galactic history of the r-process is obtained by looking at typical elements for stars of different ages. Due to the excellent observing properties of iron it is often taken as proxy for the age while the element europium is often taken as representative for the r-process. The discipline of investigating the elemental trends with time is called Galactic chemical evolution (GCE). The chemical evolution hereby depends on many known, but often also unknown inputs such as the star formation rate, the initial mass function (i.e., how probable it is to form a star with certain mass), galactic dynamics and mixing processes as well as rates and theoretically constrained nuclear yields of all relevant and chemically contributing events \citep[see, e.g.,][for reviews]{Tinsley1980,Matteucci2021}.

When interpreting observed elemental abundance trends there are some caveats to consider. Self-pollution of the atmosphere by elements synthesised in deeper regions of the star and changing the abundances of the atmosphere cannot always be ruled out \citep{Gratton2000,Denissenkov2003,Spite2005,Lagarde2012,Placco2014,Henkel2018}. The often made assumption of local thermal equilibrium (LTE) as well as the difference of 1D and 3D stellar atmospheric models can introduce artificial abundance trends \citep{Nordlander2017,Bjorgen2017,Bergemann2017,Bergemann2019,Norris2019,Gallagher2020}. Using different methods to derive elemental abundances will furthermore introduce possible systematic deviations when compiling abundances over several sources of literature \citep{Mashonkina2017a,Mashonkina2017b,Reichert2020,Minelli2021}. On top of all that, stars that contain a greater abundance in an element will more likely be displayed as a detection as well as brighter stars can be observed with better quality leaving behind and observational bias.

Having all this in mind, a careful assessment of abundance trends can be performed to get hints of the r-process element contributor(s) in the Galaxy. A directly observed contributor in galactic history are neutron star mergers (NSM, \citealt{Abbott_et_al__2017__PhysicalReviewLetters__GW170817ObservationofGravitationalWavesfromaBinaryNeutronStarInspiral,Smartt2017,Kasen2017,Pian2017,Tanvir__2017__TheAstrophysicalJournal__TheEmergenceofaLanthanideRichKilonovaFollowingtheMergerofTwoNeutronStars,Watson2019}). Therefore, all observed abundance trends have to be explained as a superposition to the r-process contribution of NSM. The precise contribution of NSM to the total r-process content of the universe is, however, still under debate. The main question that arises is: Can NSM alone explain all r-process abundance trends and peculiarities in the Universe?

If one assumes that they are the only event in the universe that synthesizes r-process elements, the observation of old, metal-poor stars with an abundance enhancement of r-process elements is challenging to explain. The neutron stars need a certain time to merge, which is given by the time two CCSNe need to explode and the following inspiral of the NSs. It has been put into question if this time can be short enough to explain the observation of metal-poor stars with r-process signature. Most studies agree that an r-process contribution with a minimum delay time of a couple of hundreds of Myrs is necessary to explain the Galactic r-process abundances \citep[e.g.,][]{Argast2004,Matteucci2014,Cescutti2015,Wehmeyer2015,Schoenrich2019,Kobayashi2020}. Therefore other mechanisms have been pointed out to explain low metallicity r-process enriched stars. While iron can be used as a proxy for time, this only holds in case of looking at stars in the same environment. Environments with smaller star forming efficiency such as dwarf galaxies will also form less efficiently iron and a lower iron content is obtained over a longer time \citep{Tinsley1980,Grebel2003,Tolstoy2009,Reichert2020}. It is therefore possible that metal-poor r-process enhanced stars in our halo are a result of migration of less massive and less efficiently star forming dwarf galaxy systems \citep{Lai2011,Ishimaru2015,Komiya2016,Ojima2018,Beniamini2018,Roederer2018,Wanajo2021,Hirai2022}. Alternatively, winds or accretion flows of pristine gas can mix hydrogen into the interstellar medium of the later forming star and therefore lowering the iron with respect to the hydrogen fraction, but not europium with respect to iron \citep{Shen2015,vandevoort2015,Wanajo2021}. However, this is unable to explain a tight europium versus iron correlation that has been found for low metallicities and that has been interpreted as similar delay time of the r-process host event compared to regular CC-SNe \citep{Skuladottir2019,Skuladottir2020,Reichert2020,Farouqi2021}. This trend is difficult and complex to model and can be, due to the many input parameters, highly degenerate \citep[e.g.,][]{vandevoort2020}. Therefore it is not fully clear if this correlation is caused by NSM with different properties from what is currently thought or if it can inevitably be only explained by an additional source of r-process elements. Additionally, GCE models point out that the currently derived delay time distribution, i.e., the probability of NSM to occur after a certain time, of $\propto t^{-1}$ \citep[e.g.,][]{Piran1992,Graur2014,Chruslinska2018} is unable to explain an observed decreasing trend of the europium over iron fraction at higher metallicities \citep{Cote2019,Simonetti2019}. This indicates either a different form of the delay time distribution \citep{DAvanzo2015,Beniamini2019,Zevin2022} or an additional source that acts at early times and fades away later \citep{Cote2019}.  

Besides the very heavy elements, lighter elements of individual stars could indicate the direct signatures of Jet-SNe as well. Stars with an unusual enhancement of zinc with respect to iron have been focus of several studies. These stars are usually found early, i.e., at low metallicities in the galactic history \citep{Cayrel2004,Nissen2011,Barbuy2015,Skuladottir2017,daSilveira2018}. The disconnected chemical evolution of iron and zinc raised the question if there is another zinc contributor in the early Universe \citep{Kobayashi2006,Duffau2017,Tsujimoto2018,Hirai2018}. The early occurrence of this discrepancy fits into the picture of an event with little delay that is able to contribute to the chemical evolution early on. While the origin of such a peculiarity in zinc abundances is still unclear, it has been pointed out that the pattern of some of these stars agree well with the theoretically calculated nucleosynthetic pattern of jet-SNe and they may therefore be direct witnesses of these events \citep{Ezzeddine2019,Yong2021}. 

All the above points indicate that the treasure trove of information given by observations is still not fully understood. Introducing an additional chemical contributor that is connected to the delay time of CCSNe would be beneficial to explain many of the previously outlined phenomena \citep[][just to name a few arguing in favor of the existence of Jet-SNe]{SiqueiraMello2014,Cescutti2015,Tsujimoto_Nishimura__2015__apjl__TheRProcessinMagnetorotationalSupernovae,Wehmeyer2015,Ji2016,Mashonkina2017,Zevin2019,Skuladottir2019,Reichert2020,Reichert2021b,Molero2021,Tsujimoto2021}. This additional event should act with little delay and be relatively rare to reproduce the observed high dispersion of r-process abundances at low metallicities. To fill this gap, there is more than one possible candidate and Jet-SNe may perfectly fit into it.

\section{Summary}
\label{Sek:Summary}

During their entire life, massive stars contribute to the synthesis of
most chemical elements in the Universe.  The various types of
explosions following the collapse of the stellar core after the
exhaustion of the initial nuclear fuel not only release the previously
produced elements into the circumstellar medium, but are also the
sites of additional nucleosynthesis as the explosion provides hot and dense enough conditions for several nuclear
processes to occur.  In the central regions, nuclei may be dissociated
only to recombine to heavy isotopes as the ejected gas expands and
cools.  All successful CCSNe form elements around and slightly beyond
the Fe group this way.  Among them, the radioactive isotope
$\AZNucleus{56}{}{Ni}$ is of particular relevance because its decay
chain powers the electromagnetic emission of the CCSNe during the first weeks after the explosion.  

Heavier elements can be synthesized by the r-process in which seed
nuclei capture several neutrons before $\beta$-decaying to stable
nuclei.  However, reaching the required high neutron-to-seed ratios is
limited to matter with very low electron fraction or high entropies.
Regular CCSNe, which explode due to neutrino heating of the ejecta in
combination with hydrodynamic instabilities, fall short of this
condition.  Rapidly rotating progenitor stars, on the other hand, may
develop explosions dominated by collimated jets.  The rotational
energy of the core acts as an additional energy source for the
explosion, which can be tapped into by strong magnetic fields.  This
explosion mechanism can overcome some of the limitations found in
detailed numerical models of regular CCSNe:
\begin{itemize}
\item The kinetic energy of the explosion can exceed the level of
  $E_{\mathrm{exp}} \sim 10^{51} \, \erg$ found for the majority of
  CCSNe population by about an order of magnitude and reach the values
  inferred for HNe.
\item Larger $^{56}$Ni masses produced by the more violent explosions can
  translate into higher luminosities.
\item Parts of the outflows are not exposed to as large neutrino
  fluxes as in regular CCSNe, which allows matter ejected from close
  to the PNS to retain its neutron richness.  Consequently, r-process
  nucleosynthesis is possible in these events.
\end{itemize}

The rapidly rotating and strongly magnetized PNS at the center of such
an explosion can continue to accelerate jets beyond the time scales of
the CCSN.  If highly relativistic outflow speeds are reached, the
observational display will be that of a PM-driven GRB. The magnetic spin-down of the PM powers a wind that can also fulfil the conditions for the r-process.

If no CCSN takes place or if continuing accretion transforms the PNS
into a BH, sufficiently rapid rotation will cause the infalling matter
to accumulate in an accretion disk around the BH, which transforms
into a collapsar.  The disk loses matter to the BH as a consequence of
the loss of angular momentum due to hydromagnetic turbulence.
Furthermore, it drives a wind from its surface and emits
(anti-)neutrinos whose annihilation can, together with the magnetic
field threading the disk or connecting it to the BH, launch a GRB jet.
During the first phase of its evolution, neutrinos are trapped in the
the disk, which is characterized by a low electron fraction.  Matter
escaping the disk is thus capable of synthesizing heavy elements via
the r-process.

The picture of nucleosynthesis in jet-driven and jet-associated
supernovae is to a large degree the result of numerical models that
have become increasingly sophisticated over the last decades.  Direct
observations of individual events have not yet been able to test the
theoretical results similarly to neutron-star mergers for which the
r-process has been confirmed.  However, the chemical history of our
Galaxy and dwarf galaxies in the local group can put constraints on
the possible contributions of different r-process sites.  They suggest that rare event that produce relatively
large amounts of heavy nuclei and have only short delay times after
the onset of star formation may have played an important role during
the early galactic evolution.  Jet-like CCSNe and long GRBs driven by
rapidly rotating stars fit this description.  Apart from the
properties of the outflows, they are rare events because the required
rotation rates are not common among massive stars.  Furthermore,
this requirement is more easily met in the early Universe because the
weaker stellar winds at low metallicity reduce the loss of angular
momentum during the stellar evolution.

To cover the remaining gaps in the understanding of this class of
explosions, further progress is desirable.  This includes gradual
methodological improvements: simulations with longer simulation times,
finer resolution, more accurate neutrino transport, {more realistic equation of states, experimentally and theoretically better constrained nuclear reaction rates and mass models,} more complete {(ideally space-based high resolution)}
observational surveys {that are based on 3D atmosphere models involving the effect of magnetic fields, non-local thermodynamic equilibrium as well as experimentally better determined atomic physics properties} and a more accurate modelling of the chemical evolution
of galaxies.  Studies that explore the
amplification of the magnetic field in the PNS or disk and multi-dimensional pre-collapse models accurately predicting the profiles of rotational
velocity and magnetic field for a large set of stellar masses and
metallicities would be of great value.  The biggest step forward would of course be a direct multi-messenger observation of such an event.

\begin{acknowledgement}
  We acknowledge the support through the grant PID2021-127495NB-I00
  funded by MCIN/AEI/10.13039/501100011033 and by the European Union,
  and the Astrophysics and High Energy Physics programme of the
  Generalitat Valenciana ASFAE/2022/026 funded by MCIN and the
  European Union NextGenerationEU (PRTR-C17.I1).
  MO acknowledges support from the Spanish
  Ministry of Science via the Ram\'on y Cajal programme
  (RYC2018-024938-I) and from the Deutsche Forschungsgemeinschaft
  (DFG, German Research Foundation) - Projektnummer 279384907 - SFB
  1245.
  MR acknowledges support from the Spanish Ministry of Science via the
  Juan de la Cierva programme (FJC2021-046688-I).
\end{acknowledgement}

\end{document}